\documentclass[aps,prb,reprint,longbibliography,superscriptaddress]{revtex4-2}

\usepackage{amsmath,amssymb,bm,braket}
\usepackage{graphicx}
\usepackage{xcolor}
\usepackage[colorlinks,linkcolor=blue,citecolor=blue,urlcolor=blue]{hyperref}
\usepackage{natbib}

\begin{document}
\title{Interference of critical dynamics associated with zero modes}

\author{Zhi-Han Zhang}
\affiliation{College of Physics, Sichuan University, 610064, Chengdu, People's Republic of China}

\author{Han-Chuan Kou}
\affiliation{School of Sciences, Southwest Petroleum University, Nanchong 637001, China}

\author{Peng Li}
\email{lipeng@scu.edu.cn}
\affiliation{College of Physics, Sichuan University, 610064, Chengdu, People's Republic of China}

\begin{abstract}
We study the interference of critical dynamics associated with zero modes (ICDZM) in the generalized Creutz ladders using closed quench paths that pass through two critical points successively. By reading out the final zero-mode transfer probability, we find rich ICDZM interference patterns dependent on the quench path. In particular, when the closed path links two topologically nontrivial phases, the ICDZM pattern may either vanish or exhibit period doubling. Within the framework of WKB analysis, this phenomenon is well clarified by the interference phase accumulated in the quench procedure. We also demonstrate that the zero-mode transfer probability can be detected by the deviation of the boundary particle number from its initial fractional value, which arises from the blending of bulk modes in the critical dynamics. As an edge defect, the zero-mode transfer probability captures both the ICDZM oscillation and the known anomalous defect production in a non-closed quench path. These results identify ICDZM and the corresponding edge defect as probes for critical dynamics associated with topological zero modes.
\end{abstract}
\maketitle


\section{Introduction}
Nonadiabatic dynamics across continuous phase transitions has become a central theme in nonequilibrium many-body physics. A natural starting point is the Kibble-Zurek (KZ) mechanism, which describes the generation of excitations when a system is driven through a critical point at a finite rate \cite{Kibble1976,Zurek1985}. In its conventional form, the KZ mechanism describes defect production with characteristic scaling law. In recent decades, it has been extensively verified and explored from both theoretical \cite{Dziarmaga2005,ZurekDornerZoller2005,Campo2010,Dziarmaga2010AdvPhys, Campo2018, Zurek2019, Campo2019, Dziarmaga2019, Campo2020,Dziarmaga2021, Kou2023} and experimental \cite{Ulm2013,Pyka2013,Navon2015Science,Keesling2019Nature,Gherardini2024PRL,JaraCosme2024PRB} perspectives and is usually taken as a standard paradigm for analyzing critical dynamics. 

On the other hand, topological zero modes stand as one of the most distinctive manifestations of topology in low-dimensional systems. The study of such modes ranges from early studies of fractionalized zero-energy states and solitons~\cite{JackiwRebbi1976,SSH1979,Creutz1999} to modern formulations of Majorana zero modes in topological superconductivity~\cite{Kitaev2001,FuKane2008,Lutchyn2010,Oreg2010,Alicea2012,Beenakker2013,Kells2014,DasSarma2015NPJQI,SatoAndo2017,Lutchyn2018NatRevMat,Yang2018,Ummode2020,Jafari2021,Sim2022,Li2022,Nie2024,Pan2025}, and further to a wide range of experimental and synthetic-platform realizations~\cite{Mourik2012Science,NadjPerge2014Science,Albrecht2016Nature,Zhang2018Nature,Wang2018Science,Hafezi2013NatPhoton,Rechtsman2013Nature,Mancini2015Science,Stuhl2015Science,Imhof2018NatPhys,Bandres2018Science,Ozawa2019RMP}. In particular, in non-interacting or effectively single-particle quantum simulators, SSH-type topological boundary states have been directly probed in momentum-space lattices and Rydberg-atom synthetic dimensions~\cite{Gadway2016,Kanungo2022}. These experiments establish the accessibility of elementary edge-state structures, while the observable studied is defined for a filled fermionic ground state. Together with related boundary modes, topological zero modes have remained a central theme in condensed-matter and quantum-simulation research~\cite{KaneMele2005,Bernevig2006Science,Konig2007Science,Chiu2016RMP,Bansil2016RMP,Goldman2016NatPhys,Aidelsburger2018CRPhys,Cooper2019RMP,Prada2020NatRevPhys,Flensberg2021NatRevMater,Jack2021NatRevPhys,Kanungo2022NatCommun,ArguelloLuengo2024CommunPhys,Yu2025PhotonicsInsights}. This naturally raises the question of how critical dynamics can be connected to many-body excitations associated with zero modes.

A natural starting point for investigating the nonequilibrium dynamics of zero modes is to employ the usual one-way quench protocol, i.e., the system is driven across just one critical point. In the Creutz ladders, topology can induce anomalous defect production beyond conventional bulk KZ scaling \cite{Bermudez2009PRL}. Related work on Majorana modes showed that the nonequilibrium dynamics of zero modes exhibits delocalization and other features not captured by bulk observables alone \cite{Bermudez2010NJP, Lee2015PRB, Li2026}. More recently, zero-mode and bulk-state transport have also been found to obey distinct nonequilibrium scaling laws in driven one-dimensional topological systems \cite{Huang2024PRB}. Recent work has further shown that topological zero modes can directly govern quantum critical dynamics, for example by inducing anomalous critical transport in disordered SSH chains \cite{Komissarov2026PRL}. However, the outcomes of the one-way quench protocol primarily reflect the generated nonadiabatic excitations and therefore cannot fully expose the information embedded in the critical dynamics.

Intriguingly, bulk interference of critical dynamics (ICD) provides a way to probe the Landau-Zener-Stueckelberg (LZS) phase accumulated during the evolution~\cite{Shevchenko2010LZS}. After two successive critical dynamics, the defect density develops oscillations along the KZ scaling law~\cite{Kou2022PRB,Zhang2025PRB}. This mechanism is closely related to interferometry.In many settings, the parameter region of interest is accessible only transiently, as in pulsed-field measurements or ultrafast pump-probe control of quantum materials~\cite{Herlach1999PulsedMagnets, Stojchevska2014HiddenState,Kohama2015FlatTopPulses, Basov2017PropertiesOnDemand,Kohama2022TimeResolvedPulsedFields, Zong2023UltrafastTechniques,Yang2023TerahertzControl}.  Notably, the period and amplitude of the interference pattern encode more information that cannot be accessed via the usual one‑way quench protocol. This suggests that the interference effect could serve as a tool to uncover the critical dynamics of topological zero‑mode depletion.

In this work, we investigate ICD associated with zero modes (ICDZM) in the generalized Creutz ladders. Protocols with closed quench paths are employed, in which a system visits two distinct phases and passes through two critical points before returning to the starting point, thereby producing ICDZM. First, we observe that closed paths can produce rich phenomena, specifically, the ICDZM pattern may vanish or exhibit period doubling when the closed path links two topologically nontrivial phases. The mechanism for these phenomena are well clarified by interference phase related to zero modes accumulated in the quench in the framework of WKB analysis. Second, we show that the edge defect due to ICDZM can be readily inferred from the boundary particle number, which is equivalent to the concept of boundary charge.

The paper is organized as follows. Section~\ref{sec:model} introduces the generalized Creutz ladders and three quench protocols. Section~\ref{sec:protocols_results} presents the bulk excitation density and zero-mode transfer probability and demonstrates rich interference patterns produced by ICD and ICDZM. Section~\ref{sec:energy_gap_period} reveals the mechanism of the interference patterns, ascribing it to the interference phase accumulated along the quench path. Section~\ref{sec:physical_observables} discusses how to measure the zero-mode transfer probability, i.e., the edge defect, through the boundary particle number deviating from its initial fractional value, which faithfully captures the ICDZM pattern. Finally, Section~\ref{sec:conclusion} summarizes the results.

\section{Generalized Creutz ladders and quench protocols}
\label{sec:model}

\subsection{Model Hamiltonian}

We consider the generalized Creutz ladders, whose real-space Hamiltonian is
\begin{equation}
\begin{split}
H= -\sum_j \Big[
&J_X e^{i\theta} a_{j+1}^{\dagger}a_j
+J_X e^{-i\theta} b_{j+1}^{\dagger}b_j
+J_Y e^{-i\phi} a_j^{\dagger}b_j  \\
&+J_D a_j^{\dagger}b_{j+1}
+J_D b_j^{\dagger}a_{j+1}
+\mathrm{H.c.}
\Big].
\end{split}
\label{eq:generalized_creutz_real}
\end{equation}
Here \(a_j^\dagger\) and \(b_j^\dagger\) create fermions on the two legs of the \(j\)th rung. The parameters \(J_X\), \(J_Y\), and \(J_D\) denote the hopping amplitudes, while \(\theta\) and \(\phi\) are the phases carried by the hopping terms. The generalized form of Eq.~\eqref{eq:generalized_creutz_real} yields the parameter space required to compare
distinct quench paths within the same model~\cite{Ning2017}.

Under the open boundary condition (OBC) that we mainly focus on, the Hamiltonian can be diagonalized as follows:
\begin{equation}
H=\sum_{j=1}^{L}E_{j,+}\,\eta_{j,+}^\dagger\eta_{j,+}
-E_{j,-}\,\eta_{j,-}^\dagger\eta_{j,-}.
\label{eq:creutz_diag_form_main}
\end{equation}
where the quasiparticle operators read
\begin{equation}
\eta_{\beta}
=
\sum_{l=1}^{L}
u_{\beta,l}^{*}a_l
+
v_{\beta,l}^{*}b_l,
\label{eq:generalized_creutz_eta_beta}
\end{equation}
with \(\beta=(j,\pm)\), \(j=1,\cdots,L\), and \(\pm\) labels the positive or negative energy levels \(E_{j,+}\) and \(-E_{j,-}\).
In Eq.~\eqref{eq:creutz_diag_form_main}, the modes are labeled in order of increasing energy value: \(0\le E_{1,\pm}\le E_{2,\pm}\le \cdots\). In the topological regime, \(\eta_{1,+}\) and \(\eta_{1,-}\) denote zero modes in the thermodynamic limit, whose energies \(E_{1,+}\) and \(-E_{1,-}\) are exponentially small in system size. At half filling, we consider two degenerate ground states that can be expressed as
\begin{equation}
|\Psi_0\rangle
=
\prod_{j=1}^{L}
\eta_{j,-}^{\dagger}|0\rangle ,
\label{eq:ground_state_main}
\end{equation}
\begin{equation}
|\Psi_0'\rangle
=
\eta_{1,+}^{\dagger}\eta_{1,-}|\Psi_0\rangle .
\label{eq:ground_state_partner_main}
\end{equation}

In the following, we restrict the model to the parameter plane
\begin{equation}
\phi=0,
\quad
J_X=J_D=K,
\label{eq:creutz_plane}
\end{equation}
and introduce the dimensionless parameter
\begin{equation}
\mu=\frac{J_Y}{2K}.
\label{eq:mu_def}
\end{equation}
For \(\mu=0\), the model reduces to the original Creutz ladders~\cite{Creutz1999}. Therefore, the present parametrization recovers the previous Creutz ladders as a special case, while allowing us to deviate from it by varying \(J_Y\) or \(\mu\).

\subsection{Phase diagram}

To better understand the quench protocols, we first establish the phase diagram of the generalized Creutz ladders by imposing periodic boundary conditions (PBC). With the Fourier transformation
\begin{equation}
a_j=\frac{1}{\sqrt L}\sum_k a_k e^{-ikj},
\quad
b_j=\frac{1}{\sqrt L}\sum_k b_k e^{-ikj},
\label{eq:fourier_transform}
\end{equation}
the Hamiltonian can be written as
\begin{equation}
H
=
\sum_k
\Psi_k^\dagger h_k(\theta,\mu)\Psi_k,
\quad
\Psi_k=(a_k,b_k)^T .
\label{eq:bloch_form}
\end{equation}
The Bloch Hamiltonian reads
\begin{equation}
h_k(\theta,\mu)
=
d_{0,k}(\theta)I
+
d_{x,k}(\mu)\sigma_x
+
d_{z,k}(\theta)\sigma_z ,
\label{eq:bloch_hamiltonian}
\end{equation}
where
\begin{equation}
\left\{
\begin{aligned}
d_{0,k}(\theta)&=-2K\cos\theta\cos k,\\
d_{x,k}(\mu)&=-2K(\cos k+\mu),\\
d_{z,k}(\theta)&=2K\sin\theta\sin k.
\end{aligned}
\right.
\label{eq:d_vector}
\end{equation}
By the transformation
\begin{equation}
    \left\{\begin{split}
        &\eta_{k,+}=u_{+}(k)a_k+v_{+}(k)b_k,\\
        &\eta_{k,-}=u_{-}(k)a_k+v_{-}(k)b_k,
    \end{split}\right.
    \label{eq:k_diag_transformation}
\end{equation}
the Hamiltonian can be diagonalized as
\begin{equation}
    H=\sum_k \omega_{+}(k)\eta_{k,+}^\dagger\eta_{k,+}
    +\omega_{-}(k)\eta_{k,-}^\dagger\eta_{k,-}.
    \label{eq:diag_k_hamiltonian}
\end{equation}
where \(\omega_\pm(k)=d_{0,k}\pm2K\Lambda_k(\theta,\mu)\) and
\begin{equation}
\Lambda_k(\theta,\mu)
=
\sqrt{(\cos k+\mu)^2+\sin^2\theta\,\sin^2k}.
\label{eq:Lambda_general}
\end{equation}

\begin{figure}[t]
    \centering
    \includegraphics[width=1\linewidth]{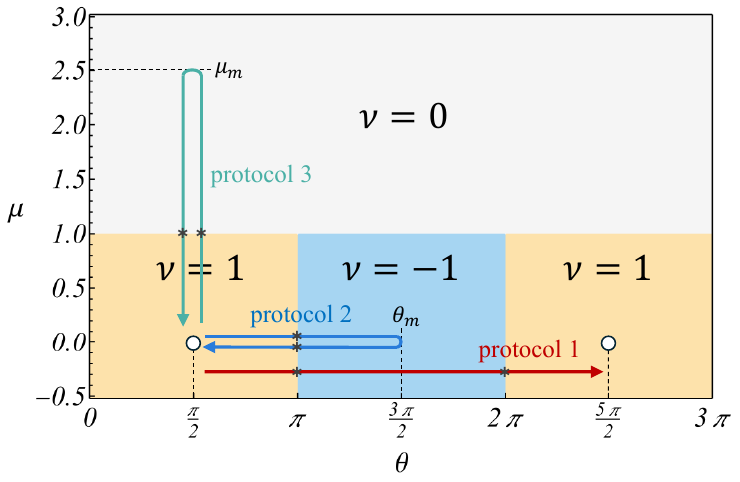}
    \caption{Phase diagram and quench protocols of the generalized Creutz ladders with \(\phi=0\) and \(J_X=J_D=K\). The vertical axis is \(\mu=J_Y/(2K)\). The colored regions are labeled by the winding number \(\nu\). The equivalent solvable endpoints \(\theta=\pi/2, \mu=0\) and \(\theta=5\pi/2, \mu=0\) host OBC zero modes exactly localized on boundary rungs. In protocol 1, \(\theta\) is varied from \(\pi/2\) to \(5\pi/2\) at fixed \(\mu=0\). In protocol 2, \(\theta\) is varied from \(\pi/2\) to the turning point \(\theta_m=3\pi/2\) and then back to \(\pi/2\) at fixed \(\mu=0\). In protocol 3, \(\mu\) is varied from \(0\) to the turning point \(\mu_m=2.5\) and then back to \(0\) at fixed \(\theta=\pi/2\). Small offsets of the arrows are used only for visibility. Asterisks (*) denote the critical points where the paths cross the phase boundaries.}
    \label{fig:GCreutz_phase}
\end{figure}

The identity term \(d_{0,k}I\) shifts both branches equally and does not affect the Bloch eigenvectors. The topological sectors are therefore determined by the vector \((d_{x,k},d_{z,k})\) in the \(x\)-\(z\) plane. We use the signed winding number
\begin{equation}
\nu
=
-\frac{1}{2\pi}
\int_{\rm BZ}dk\,
\frac{
d_{x,k}\partial_k d_{z,k}
-
d_{z,k}\partial_k d_{x,k}
}{
d_{x,k}^2+d_{z,k}^2
}.
\label{eq:winding_def}
\end{equation}
For \(\sin\theta\neq0\), the result is
\begin{equation}
\nu=
\begin{cases}
\mathrm{sgn}(\sin\theta), & |\mu|<1,\\[0.5ex]
0, & |\mu|>1.
\end{cases}
\label{eq:winding_general}
\end{equation}
Thus \(|\mu|=1\) separates the topologically nontrivial sectors from the trivial sector, while \(\theta=0,\pm\pi\) separate the two nontrivial sectors with opposite winding numbers. The resulting phase diagram is shown in Fig.~\ref{fig:GCreutz_phase}.

\subsection{Quench protocols}
\label{subsec:quench_protocols}
We consider three quench protocols, as shown in Fig.~\ref{fig:GCreutz_phase}.
All of them start from the point \((\theta_i,\mu_i)=\left(\pi/2,0\right)\),
where the OBC zero modes are exactly localized on the boundary rungs:
\begin{equation}
    \eta_{1,-}^{\dagger}
    =
    \frac{1}{\sqrt2}
    \left(
    a_1^\dagger-i b_1^\dagger
    \right).
\label{eq:phys_left_zero_mode_form}
\end{equation}

Protocols 1 and 2 are generated by varying \(\theta\) at fixed \(J_Y=0\), namely \(\mu=0\). In this parameter family, the vertical hopping term is absent, and the OBC Hamiltonian has the hidden particle-hole symmetry and hidden chiral symmetry of the rungless Creutz ladder~\cite{Zurita2021Quantum}. 

In protocol 1, \(\mu=0\) is fixed, while \(\theta\) is varied from \(\theta_i=\pi/2\) to the equivalent point \(\theta_f=5\pi/2\):
\begin{equation}
\theta(t)=
\begin{cases}
\theta_m+\dfrac{t}{\tau_Q},
& t_i\le t\le 0,\\[1.0ex]
\theta_m+\dfrac{t}{\tau_Q'},
& 0\le t\le t_f,
\end{cases}
\label{eq:protocol1_theta_main}
\end{equation}
with the intermediate point \(\theta_m=3\pi/2\). This protocol crosses the critical points \(\theta=\pi\) and \(\theta=2\pi\), and enters the sector with \(\nu=-1\) between the two crossings.

In protocol 2, \(\mu=0\) is fixed again, while \(\theta\) is varied from \(\theta_i=\pi/2\) to \(\theta_m=3\pi/2\) and then back to \(\theta_f=\pi/2\):
\begin{equation}
\theta(t)=
\begin{cases}
\theta_m+\dfrac{t}{\tau_Q},
& t_i\le t\le 0,\\[1.0ex]
\theta_m-\dfrac{t}{\tau_Q'},
& 0\le t\le t_f .
\end{cases}
\label{eq:protocol2_theta_main}
\end{equation}
This protocol crosses the same transition point \(\theta=\pi\) twice.

Protocol 3 is generated by varying \(J_Y\), or equivalently \(\mu=J_Y/(2K)\), at fixed \(\theta=\pi/2\). At this value of \(\theta\), the Creutz ladder has the standard chiral and particle-hole symmetries represented in the two-leg space \cite{Roy2023PRB,Chiu2016RMP}. The system is initially prepared at \(\mu_i=0\). It is then driven to \(\mu_m=2.5\) and subsequently brought back to \(\mu_f=0\):
\begin{equation}
\mu(t)=
\begin{cases}
\mu_m+\dfrac{t}{\tau_Q},
& t_i\le t\le 0,\\[1.0ex]
\mu_m-\dfrac{t}{\tau_Q'},
& 0\le t\le t_f .
\end{cases}
\label{eq:protocol3_mu_main}
\end{equation}
Thus the system enters the trivial region \(\nu=0\) and crosses the critical point \(\mu=1\) twice.

During these quench protocols, the Hamiltonian becomes time dependent through a control parameter \(\lambda(t)\), which is taken to be either \(\mu(t)\) or \(\theta(t)\). We write
\begin{equation}
    H(t)\equiv H[\lambda(t)] ,
\end{equation}
and the initial ground state evolves as
\begin{equation}
    |\psi(t)\rangle
    =
    U(t,t_i)|\Psi_0\rangle,
\label{eq:time_evolution_main}
\end{equation}
with the time-evolution operator $U(t,t_0) = \mathcal{T} \exp\left(-i\int_{t_0}^{t} H(\tau) d\tau\right)$, where $\mathcal{T}$ is the time-ordering operator. Equivalently, the evolution obeys Heisenberg equations:
\begin{equation}
\begin{aligned}
    i\frac{d}{dt} \tilde{a}_j(t)
    =
    [H(t),\tilde{a}_{j}^{\dagger}(t)],\quad i\frac{d}{dt} \tilde{b}_j(t)
    =
    [H(t),\tilde{b}_{j}^{\dagger}(t)],
\label{eq:heisenberg_eta_main}
\end{aligned}
\end{equation}
with \(\tilde{a}_j(t)=U^\dagger(t,t_i) a_jU(t,t_i)\) and \(\tilde{b}_j(t)=U^\dagger(t,t_i) b_jU(t,t_i)\). Since the Hamiltonian is quadratic and conserves the particle number, Eq.~\eqref{eq:heisenberg_eta_main} is closed in the single-particle space.

At each time, the instantaneous eigenmode pair is denoted b \(\eta_{j,\pm}[\lambda(t)]\). The corresponding occupation operators are defined as
\begin{equation}
    \hat n_{j,\pm}(t)
    =
    \eta_{j,\pm}^{\dagger}[\lambda(t)]
    \eta_{j,\pm}[\lambda(t)],
\end{equation}
and the corresponding expectation values are
\begin{equation}
    n_{j,\pm}(t)
    =
    \langle\psi(t)|\hat n_{j,\pm}(t)|\psi(t)\rangle .
\label{eq:instantaneous_occupation_main}
\end{equation}
For the half-filled initial state in Eq.~\eqref{eq:ground_state_main}, the particle-hole symmetry of the single-particle Hamiltonian relates the two members of each instantaneous pair \((j,+)\) and \((j,-)\). As a result, the pair occupation satisfies
\begin{equation}
    n_{j,+}(t)+n_{j,-}(t)=1 .
\label{eq:mode_pair_sum_main}
\end{equation}
The real-space evolution equation for the OBC Hamiltonian and the proof of Eq.~\eqref{eq:mode_pair_sum_main} are given in Secs.~S1 and S2 of the Supplemental Material~\cite{SM}, respectively. At the topologically nontrivial phase, the \(j=1\) pair is the zero-mode pair. In the trivial region of protocol 3, the same notation denotes the instantaneous lowest-energy pair.

Unless otherwise specified, the numerical calculations below are performed with \(L=500\), \(K=1\), and \(R=1\).

\section{Zero-mode transfer probability and interference patterns}
\label{sec:protocols_results}

In this section, we introduce the zero-mode transfer probability and compare it with the bulk excitation density under the three protocols defined in Sec.~\ref{sec:model}.

\begin{figure*}[t]
    \centering
    \includegraphics[width=1\textwidth]{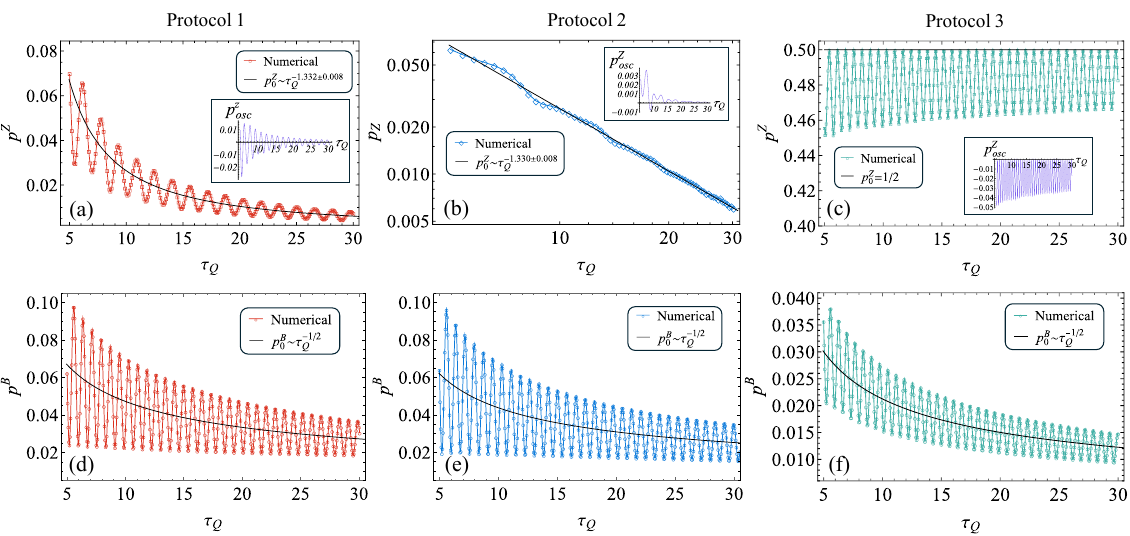}
    \caption{
        Zero-mode transfer probability \(p^{\mathrm Z}\) and bulk excitation density \(p^{\mathrm B}\) for the three quench protocols. (a)-(c) Zero-mode transfer probability \(p^{\mathrm Z}\). The black curves denote the smooth component \(p_0^{\mathrm Z}\), and the insets show the oscillatory part \(p_{\rm osc}^{\mathrm Z}\). (d)-(f) Bulk excitation density \(p^{\mathrm B}\). The black curves show the KZ scaling \(p_0^{\mathrm B}\sim\tau_Q^{-1/2}\). Panels (a,d), (b,e), (c,f) correspond to protocols 1, 2, and 3, respectively. The numerical parameters are \(L=500\), \(K=1\), and \(R=1\).
    }
    \label{fig:ICDZM_patterns_main}
\end{figure*}

The initial state is prepared as the half-filled ground state in Eq.~\eqref{eq:ground_state_main}. At the initial point, the selected occupied zero mode is \(\eta_{1,-}\). After the quench, the final occupation of the selected zero mode is
\begin{equation}
    n_{1,-}(t_f)
    =
    \langle\psi(t_f)|
    \eta_{1,-}^{\dagger}(\lambda_f)
    \eta_{1,-}(\lambda_f)
    |\psi(t_f)\rangle ,
\label{eq:n1minus_final_main}
\end{equation}
where \(\lambda_f\) denotes the final value of the driven parameter. We define the zero-mode transfer probability as
\begin{equation}
    p^{\mathrm Z}(t_f)
    =
    1-n_{1,-}(t_f).
\label{eq:pZ_main_def}
\end{equation}
Using the pair-occupation relation in Eq.~\eqref{eq:mode_pair_sum_main}, the final zero-mode pair satisfies
\begin{equation}
    n_{1,+}(t_f)+n_{1,-}(t_f)=1 .
\label{eq:zero_pair_sum_secIII}
\end{equation}
Therefore, we have \(p^{\mathrm Z}(t_f)=n_{1,+}(t_f)\).  Thus \(p^{\mathrm Z}\) measures the probability transferred from the selected occupied zero mode \(\eta_{1,-}^\dagger\) to its particle-hole partner \(\eta_{1,+}^\dagger\). As a comparison, we also calculate the bulk excitation density,
\begin{equation}
    p^{\mathrm B}(t_f)
    =
    \frac{1}{L}
    \sum_{j=2}^{L}
    \langle\psi(t_f)|
    \eta_{j,+}^{\dagger}(\lambda_f)
    \eta_{j,+}(\lambda_f)
    |\psi(t_f)\rangle .
\label{eq:pB_main_def}
\end{equation}

For a protocol containing two successive critical dynamics, the bulk excitation density can be written as
\begin{equation}
    p^{\mathrm B}(\tau_Q)
    =
    p_0^{\mathrm B}(\tau_Q)
    +
    p_{\rm osc}^{\mathrm B}(\tau_Q),
\label{eq:pB_decomposition_main}
\end{equation}
where the smooth part follows the KZ scaling law, \(p_0^{\mathrm B}(\tau_Q)\sim \tau_Q^{-1/2}\) and the oscillatory part gives the ordinary ICD signal \cite{Kou2022PRB}. When the oscillatory part is visible, it can be written as
\begin{equation}
    p_{\rm osc}^{\mathrm B}(\tau_Q)
    \sim
    \cos\left(
    \frac{2\pi}{T_{\mathrm B}}\tau_Q+\varphi_{\mathrm B}
    \right),
\label{eq:pB_osc_form_main}
\end{equation}
where \(T_{\mathrm B}\) is the bulk ICD period. Similarly, the zero-mode transfer probability is decomposed as
\begin{equation}
    p^{\mathrm Z}(\tau_Q)
    =
    p_0^{\mathrm Z}(\tau_Q)
    +
    p_{\rm osc}^{\mathrm Z}(\tau_Q).
\label{eq:pZ_decomposition_main}
\end{equation}
Here \(p_0^{\mathrm Z}\) denotes the smooth part attributed to an anomalous scaling law, and \(p_{\rm osc}^{\mathrm Z}\) denotes the oscillatory part associated with ICDZM. When this contribution gives a clear oscillation, we write
\begin{equation}
    p_{\rm osc}^{\mathrm Z}(\tau_Q)
    \sim
    \cos\left(
    \frac{2\pi}{T_{\mathrm Z}}\tau_Q+\varphi_{\mathrm Z}
    \right),
\label{eq:pZ_osc_form_main}
\end{equation}
where \(T_{\mathrm Z}\) is the ICDZM period.

We first consider protocol 1, where \(\theta\) is varied at fixed \(\mu=0\) from \(\pi/2\) to the equivalent point \(5\pi/2\). This protocol crosses two different critical lines, \(\theta=\pi\) and \(\theta=2\pi\), and passes through the sector with \(\nu=-1\). As shown in Fig.~\ref{fig:ICDZM_patterns_main}(a), the zero-mode transfer probability remains much smaller than unity. According to Eq.~\eqref{eq:zero_pair_sum_secIII}, this means that only a small part of the selected zero-mode occupation is transferred to its particle-hole partner.

The smooth background follows an anomalous power law,
\begin{equation}
    p_0^{\mathrm Z}
    \sim
    \tau_Q^{-\kappa_{1}},
\label{eq:protocol1_p0Z_scaling}
\end{equation}
with \(\kappa_{1}=1.332\pm0.008\). This exponent is close to the anomalous zero-mode response reported for one-way quench dynamics in the Creutz ladder~\cite{Bermudez2009PRL}. On top of this smooth background, \(p_{\rm osc}^{\mathrm Z}\) gives a clear oscillatory part. Comparing Fig.~\ref{fig:ICDZM_patterns_main}(a) with Fig.~\ref{fig:ICDZM_patterns_main}(d), we find
\begin{equation}
    T_{\mathrm Z}
    \simeq
    2T_{\mathrm B}.
\label{eq:protocol1_period_relation_secIII}
\end{equation}
Thus, in protocol 1, the zero-mode transfer probability does not reproduce the ordinary bulk ICD period. Instead, it exhibits period doubling relative to the bulk excitation density.

We next consider protocol 2, where \(\theta\) is varied at fixed \(\mu=0\) from \(\pi/2\) to \(3\pi/2\) and then back to \(\pi/2\). This protocol crosses the same critical line \(\theta=\pi\) twice. As shown in Fig.~\ref{fig:ICDZM_patterns_main}(b), the zero-mode transfer probability again remains small and is mainly described by a power-law decay,
\begin{equation}
    p_0^{\mathrm Z}
    \sim
    \tau_Q^{-\kappa_2},
\label{eq:protocol2_p0Z_scaling}
\end{equation}
with \(\kappa_2=1.330\pm0.008\). The pair relation in Eq.~\eqref{eq:zero_pair_sum_secIII} still applies, so this small value means that the selected final zero mode remains almost occupied. However, the inset of Fig.~\ref{fig:ICDZM_patterns_main}(b) shows that \(p_{\rm osc}^{\mathrm Z}\) is strongly suppressed. Therefore no stable zero-mode ICDZM period is extracted for protocol 2, even though the bulk excitation density in Fig.~\ref{fig:ICDZM_patterns_main}(e) still shows an ordinary ICD oscillation around the KZ background.

This comparison between protocols 1 and 2 shows that the pair-occupation relation fixes the final redistribution inside the zero-mode pair, but it does not determine the visibility of ICDZM. A visible ICDZM oscillation also requires a sizable oscillatory part in \(p^{\mathrm Z}\).

Finally, we consider protocol 3, where \(\mu\) is varied at fixed \(\theta=\pi/2\). The system enters the trivial region with \(\nu=0\) and then returns to the initial point. In the trivial region, the instantaneous \(j=1\) pair is the lowest-energy pair, but it is not a boundary zero-mode pair. As shown in Fig.~\ref{fig:ICDZM_patterns_main}(c), the zero-mode transfer probability oscillates below the constant smooth component
\begin{equation}
    p_0^{\mathrm Z}
    \simeq
    \frac{1}{2}.
\label{eq:protocol3_p0Z_half}
\end{equation}
Using Eq.~\eqref{eq:zero_pair_sum_secIII}, this result means
\begin{equation}
    n_{1,+}(t_f)
    \simeq
    n_{1,-}(t_f)
    \simeq
    \frac{1}{2},
\label{eq:protocol3_balanced_pair}
\end{equation}
up to the oscillatory correction. Therefore, protocol 3 produces an approximately balanced occupation of the two final zero-mode partners. The total occupation of the final zero-mode pair remains equal to one, as required by Eq.~\eqref{eq:zero_pair_sum_secIII}. This balanced value is reminiscent of the \(1/2\) contribution reported in Kitaev chain, where the class-D particle-hole symmetry relates the positive- and negative-energy partners \cite{Bermudez2010NJP,Lee2015PRB}. 

The period of the oscillatory part in Fig.~\ref{fig:ICDZM_patterns_main}(c) follows the reference period extracted from the bulk excitation density in Fig.~\ref{fig:ICDZM_patterns_main}(f),
\begin{equation}
    T_{\mathrm Z}
    \simeq
    T_{\mathrm B}.
\label{eq:protocol3_period_relation_secIII}
\end{equation}
Thus protocol 3 differs from protocols 1 and 2 in two aspects. First, the smooth zero-mode transfer background is close to \(1/2\), instead of following the anomalous decay in Eqs.~\eqref{eq:protocol1_p0Z_scaling} and \eqref{eq:protocol2_p0Z_scaling}. Second, the oscillation period follows the ordinary bulk ICD period.

The three protocols therefore give three different zero-mode transfer patterns. In protocol 1, the transfer probability is small but contains a visible period-doubled ICDZM oscillation. In protocol 2, the transfer probability is also small, but the oscillatory part is strongly suppressed. In protocol 3, the transfer probability is centered near \(1/2\), corresponding to an almost balanced occupation of the final zero-mode pair, and its oscillation follows the bulk ICD period. These results show that \(p^{\mathrm Z}\) is determined not only by the occurrence of two critical dynamics, but also by the phase visited during the quench and by the oscillatory part retained in the final zero-mode pair.

\section{Interference phase accumulated during the quench path}
\label{sec:energy_gap_period}

In this section, we explain the protocol-dependent interference periods observed in Sec.~\ref{sec:protocols_results}. For a protocol containing two successive critical dynamics, the final occupation contains an interference term determined by the relative phase accumulated between the two critical dynamics. This is the similar physical structure as LZS interferometry, where two nonadiabatic passages split and recombine the amplitudes on two instantaneous branches~\cite{Shevchenko2010LZS}.

\subsection{Quench dynamics by varying \(\theta\) and period doubling phenomenon}
\label{subsec:nontrivial_phase_secIV}

We first consider protocols 1 and 2, where \(\theta\) is varied at fixed \(\mu=0\). For PBC Hamiltonian in Eq.~\eqref{eq:bloch_form}, the Heisenberg equation reads
\begin{equation}
    i\frac{d}{dt}\Psi_k(t)
    =
    h_k[\theta(t)]\Psi_k(t),
\label{eq:theta_heisenberg_secIV}
\end{equation}
with 
\begin{equation}
        \quad
    \Psi_k(t)=
    \begin{pmatrix}
    a_k(t)\\
    b_k(t)
    \end{pmatrix}=
    U^\dagger(t,t_i)\begin{pmatrix}
    a_k\\
    b_k
    \end{pmatrix}U(t,t_i).
\end{equation}
Substituting \(\mu=0\) into Eq.~\eqref{eq:bloch_hamiltonian}, the Bloch Hamiltonian becomes
\begin{equation}
    h_k(\theta)
    =
    d_{0,k}(\theta)I
    +
    d_{x,k}\sigma_x
    +
    d_{z,k}(\theta)\sigma_z ,
\label{eq:theta_hk_secIV}
\end{equation}
where
\begin{equation}
\left\{
\begin{aligned}
d_{0,k}(\theta)&=-2K\cos\theta\cos k,\\
d_{x,k}&=-2K\cos k,\\
d_{z,k}(\theta)&=2K\sin\theta\sin k .
\end{aligned}
\right.
\label{eq:theta_d_secIV}
\end{equation}
The term \(d_{0,k}I\) only contributes a common phase. We remove it by writing
\begin{equation}
    \Psi_k(t)
    =
    e^{-i\zeta_k(t)}
    \widetilde{\Psi}_k(t),
    \quad
    \zeta_k(t)=\int_{t_i}^{t}d_{0,k}(t')\,dt' .
\label{eq:theta_common_phase_secIV}
\end{equation}
Then the reduced spinor satisfies
\begin{equation}
    i\partial_t\widetilde{\Psi}_k(t)
    =
    \left[
    d_{x,k}\sigma_x+d_{z,k}(\theta)\sigma_z
    \right]\widetilde{\Psi}_k(t).
\label{eq:theta_reduced_eom_secIV}
\end{equation}
Writing
\begin{equation}
    \widetilde{\Psi}_k(t)
    =
    \begin{pmatrix}
    x_k(t)\\
    y_k(t)
    \end{pmatrix},
\end{equation}
one obtains
\begin{equation}
\begin{aligned}
    i\dot x_k&=d_{z,k}(\theta)x_k+d_{x,k}y_k,\\
    i\dot y_k&=d_{x,k}x_k-d_{z,k}(\theta)y_k .
\end{aligned}
\label{eq:theta_first_order_secIV}
\end{equation}
Eliminating \(y_k\), we have
\begin{equation}
    \ddot x_k+
    \left[
    d_{x,k}^2+d_{z,k}^2(\theta)+i\dot d_{z,k}(\theta)
    \right]x_k=0 .
\label{eq:theta_second_time_secIV}
\end{equation}

For a linear quench, we write
\begin{equation}
    \frac{d\theta}{dt}
    =
    s_\theta\frac{1}{\tau_s},
\label{eq:theta_linear_segment_secIV}
\end{equation}
where \(s_{\theta}=\pm1\) specifies the quench direction, and \(\tau_s=\tau_Q\) or \(\tau_Q'\). Using \(\theta\) as the evolution variable, Eq.~\eqref{eq:theta_second_time_secIV} becomes
\begin{equation}
\begin{split}
\frac{d^2x_k}{d\theta^2}
+
\Big[
&4K^2\tau_s^2
\left(\cos^2 k+\sin^2 k\sin^2\theta\right)
\\
&+
i\,s_\theta\,2K\tau_s\sin k\cos\theta
\Big]x_k=0 .
\end{split}
\label{eq:theta_second_order_secIV}
\end{equation}
Introducing the half-angle variable \(\varphi=\theta/2\), this equation takes the Whittaker-Hill-type form
\begin{equation}
    \frac{d^2X_k}{d\varphi^2}
    +
    \left[
      A_k+B_k\cos2\varphi+C_k\cos4\varphi
    \right]X_k=0,
\label{eq:theta_WH_secIV}
\end{equation}
where \(X_k(\varphi)=x_k(2\varphi)\), and
\begin{equation}
\left\{
\begin{aligned}
      A_k&=8K^2\tau_s^2(1+\cos^2 k),\\
      B_k&=8i\,s_\theta K\tau_s\sin k,\\
      C_k&=-8K^2\tau_s^2\sin^2 k .
\end{aligned}
\right.
\label{eq:theta_WH_coeff_secIV}
\end{equation}

Within the WKB approximation derived in Sec.~S3 of the
Supplemental Material~\cite{SM}, the oscillatory part of the bulk excitation density takes the form
\begin{equation}
    p_{\rm osc}^{\mathrm B}
    \sim
    \cos\left(
    \Delta\varphi_{\mathrm B}
    \right).
\label{eq:theta_bulk_osc_phase_relation_secIV}
\end{equation}
The interference phase for bulk \(\Delta\varphi_{\mathrm B}\) is accumulated between the two critical dynamics. The part before the first critical dynamics and the part after the second one change only the overall phase or the final phase offset, and do not enter the interference phase.

For protocol 1, the two critical dynamics occur at \(\theta=\pi\) and \(\theta=2\pi\). Therefore,
\begin{equation}
\begin{split}
\Delta\varphi_{\mathrm B}^{(1)}
=
&\tau_Q
\int_{\pi}^{\theta_m}
\Delta_{\mathrm B}^{(\theta)}(\theta)\,d\theta
\\
&+
R\tau_Q
\int_{\theta_m}^{2\pi}
\Delta_{\mathrm B}^{(\theta)}(\theta)\,d\theta
+
\delta .
\end{split}
\label{eq:theta_protocol1_bulk_phase_secIV}
\end{equation}
For protocol 2, the same critical line \(\theta=\pi\) is crossed twice. The interference phase is
\begin{equation}
    \Delta\varphi_{\mathrm B}^{(2)}
    =
    (1+R)\tau_Q
    \int_{\pi}^{\theta_m}
    \Delta_{\mathrm B}^{(\theta)}(\theta)\,d\theta
    +
    \delta .
\label{eq:theta_protocol2_bulk_phase_secIV}
\end{equation}
Here \(\delta\) denotes local connection phases and other terms that do not grow linearly with \(\tau_Q\). The oscillation period is determined by requiring the interference phase to change by \(2\pi\),
\begin{equation}
    T_{\mathrm B}
    =
    \frac{2\pi}
    {
    \partial_{\tau_Q}\Delta\varphi_{\mathrm B}
    }.
\label{eq:theta_period_derivative_def_secIV}
\end{equation}
Using Eq.~\eqref{eq:theta_protocol1_bulk_phase_secIV}, the bulk period of protocol 1 can be written directly in terms of the integrated gap as
\begin{equation}
    T_{\mathrm B}^{(1)}
    =
    \frac{2\pi}
    {
    \int_{\pi}^{\theta_m}
    \Delta_{\mathrm B}(\theta)\,d\theta
    +
    R
    \int_{\theta_m}^{2\pi}
    \Delta_{\mathrm B}(\theta)\,d\theta
    }.
\label{eq:theta_protocol1_period_gap_relation_secIV}
\end{equation}
Similarly, using Eq.~\eqref{eq:theta_protocol2_bulk_phase_secIV}, the period of protocol 2 is
\begin{equation}
    T_{\mathrm B}^{(2)}
    =
    \frac{2\pi}
    {
    (1+R)
    \int_{\pi}^{\theta_m}
    \Delta_{\mathrm B}(\theta)\,d\theta
    }.
\label{eq:theta_protocol2_period_gap_relation_secIV}
\end{equation}
Equations~\eqref{eq:theta_protocol1_period_gap_relation_secIV} and \eqref{eq:theta_protocol2_period_gap_relation_secIV} show that the interference period is fixed by the energy gap integrated in the interval between the two critical dynamics. For both two quench protocol, the energy gap can be described as 
\begin{equation}
    \Delta_{\mathrm B}(\theta)
    =
    4K\Lambda_{\pi/2}(\theta,0)
    =
    4K|\sin\theta|,
\label{eq:theta_bulk_gap_secIV}
\end{equation}
which leads to 
\begin{equation}
    T_{\mathrm B}^{(1)}
    =T_{\rm B}^{(2)}=
    \frac{\pi}{4K},
\label{eq:theta_TB_R1_secIV}
\end{equation}
with \(R=1\). This period is the ICD period for bulk in Figs.~\ref{fig:ICDZM_patterns_main}(d) and \ref{fig:ICDZM_patterns_main}(e).

According to the direct relation between the energy gap and the oscillation period in Eq.~\eqref{eq:theta_period_derivative_def_secIV}, the period doubling can be traced to the energy gap accumulated in the topologically nontrivial region. We now compare the energy gap in PBC \(\Delta_{\mathrm B}\) with the energy gap \(\Delta_{\mathrm Z}\) in OBC.
 
In the OBC Hamiltonian, the lowest excited states with zero modes are \(\eta_{2,+}^{\dagger}\eta_{1,-}|\Psi_0\rangle\) and \(\eta_{1,+}^{\dagger}\eta_{2,-}|\Psi_0\rangle\), 
with corresponding energy gap
\begin{equation}
    \Delta_{\mathrm Z}
    =
    E_{2,+}+E_{1,-}
    =
    E_{2,-}+E_{1,+},
\label{eq:OBC_zero_gap_secIV}
\end{equation}
where the second equality follows from the particle-hole symmetry of the
spectrum. By contrast, the bulk particle-hole excitation
\begin{equation}
    \eta_{2,+}^{\dagger}\eta_{2,-}|\Psi_0\rangle
\end{equation}
has the energy gap
\begin{equation}
    \Delta_{\mathrm B}
    =
    E_{2,+}+E_{2,-},
\label{eq:OBC_bulk_gap_secIV}
\end{equation}
which is same of that for PBC in the thermodynamic limit. In the topologically nontrivial region, \(E_{1,+}\) and \(E_{1,-}\) are exponentially close to zero, so that
\begin{equation}
    \Delta_{\mathrm Z}
    =
    \frac{1}{2}\Delta_{\mathrm B}.
\label{eq:theta_gap_half_secIV}
\end{equation}
\begin{figure}[t]
    \centering
    \includegraphics[width=0.88\linewidth]{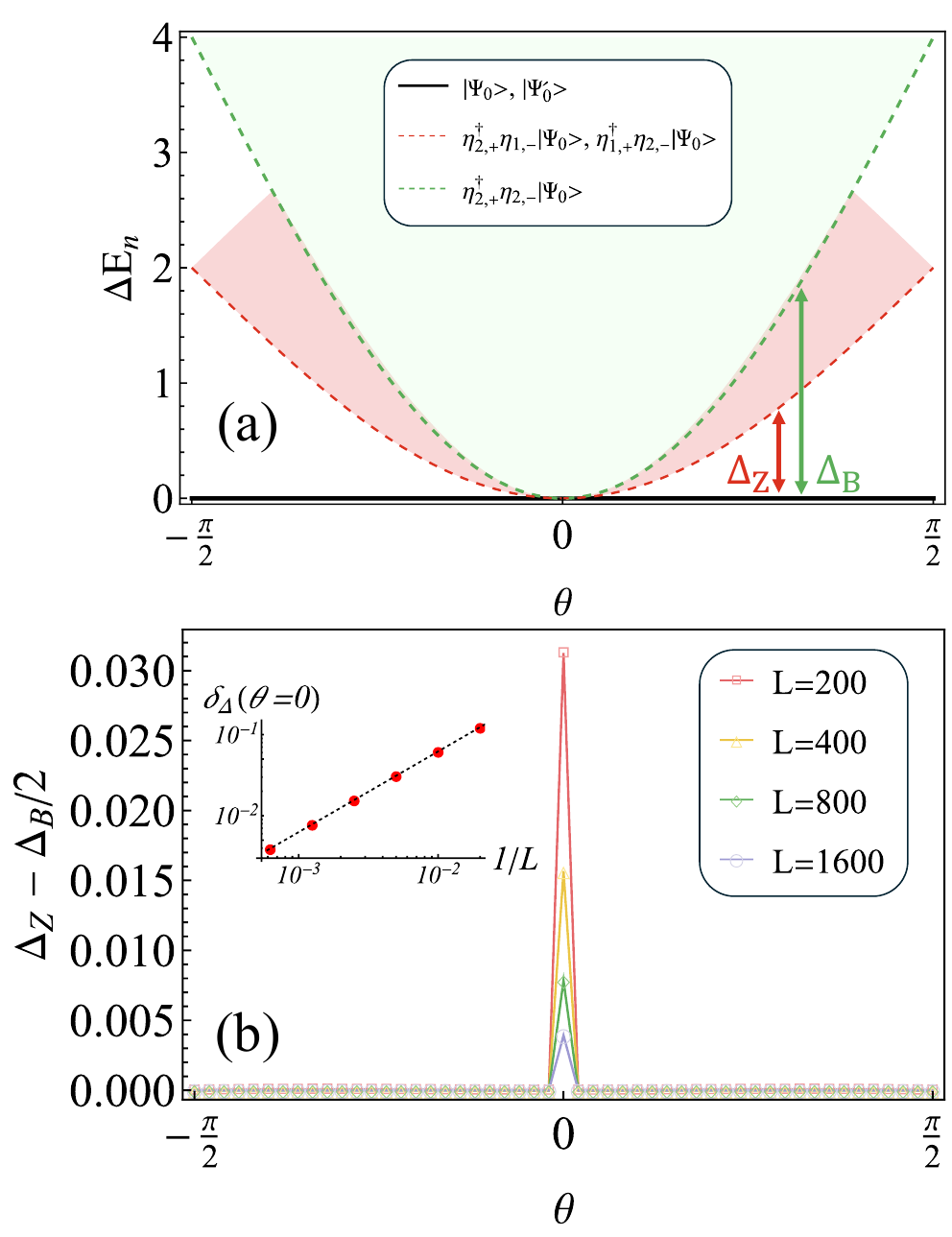}
    \caption{
    Relation between the OBC energy gap and the PBC bulk gap. (a) Many-body energy level structure. The red dashed line denotes the energy of the two degenerate excited states, \(\eta_{2,+}^{\dagger}\eta_{1,-}|\Psi_0\rangle\) and \(\eta_{1,+}^{\dagger}\eta_{2,-}|\Psi_0\rangle\), which mark the lowest energy bound of excitations involving zero modes, namely the OBC energy gap\(\Delta_{\mathrm Z}=E_{2,+}+E_{1,-}\simeq E_{2,+}\).  The green dashed line corresponds to the energy of the excited state, \(\eta_{2,+}^{\dagger}\eta_{2,-}|\Psi_0\rangle\), with bulk gap \(\Delta_{\mathrm B}=E_{2,+}+E_{2,-}= 2E_{2,+}\). The light red region marks excitations with zero modes, while the light green region marks bulk many-body excitations under PBC.
    (b) Difference between the OBC energy gap and half the PBC bulk gap, \(\Delta_{\mathrm Z} -\Delta_{\mathrm B}/2\), as a function of the quench parameter \(\theta\). The inset shows the finite-size scaling of this difference at the critical point.
    }
    \label{fig:theta_gap_compare_secIV}
\end{figure}
Figure~\ref{fig:theta_gap_compare_secIV} (b) gives the energy gap in OBC \(\Delta_Z\) one-half of the PBC bulk gap \(\Delta_B\), which verifies that the OBC energy gap is half of PBC bulk gap in Eq.~\eqref{eq:theta_gap_half_secIV}. Since the interference phase is obtained by integrating the relevant energy scale between the two critical dynamics, Eq.~\eqref{eq:theta_gap_half_secIV} gives
\begin{equation}
    \Delta\varphi_{\mathrm Z}
    \simeq
    \frac{1}{2}\Delta\varphi_{\mathrm B}
    +
    {\rm const.}
\label{eq:theta_phase_half_secIV}
\end{equation}
The constant offset does not affect the period. Therefore,
\begin{equation}
    T_{\mathrm Z}
    \simeq
    2T_{\mathrm B}.
\label{eq:theta_TZ_double_secIV}
\end{equation}

 For protocol 2, the same phase interval also lies in the topologically nontrivial region, and the  is available. However, as shown in the inset of Fig.~\ref{fig:ICDZM_patterns_main}(b), \(p_{\rm osc}^{\mathrm Z}\) is strongly suppressed. Therefore, the possible period-doubled phase is not resolved as a stable oscillation in the final zero-mode transfer probability. 

\subsection{Quench dynamics by varying \(mu\) through the topologically trivial region}
\label{subsec:trivial_phase_secIV}

We now consider protocol 3, where \(\mu\) is varied at fixed \(\theta=\pi/2\). For PBC Hamiltonian in Eq.~\eqref{eq:bloch_form}, the Heisenberg equation reads
\begin{equation}
    i\frac{d}{dt}\Psi_k(t)
    =
    h_k[\mu(t)]\Psi_k(t),
\label{eq:mu_heisenberg_secIV}
\end{equation}
Substituting \(\theta=\pi/2\) into Eq.~\eqref{eq:bloch_hamiltonian}, one obtains
\begin{equation}
    h_k(\mu)
    =
    -2K(\mu+\cos k)\sigma_x
    +
    2K\sin k\,\sigma_z .
\label{eq:mu_hk_secIV}
\end{equation}
The identity term vanishes in this protocol.

To put the time-dependent term on \(\sigma_z\), we perform the time-independent rotation
\begin{equation}
    \widetilde{\Psi}_k(t)
    =
    U_y\Psi_k(t)
    =
    \begin{pmatrix}
    u_k(t)\\
    v_k(t)
    \end{pmatrix},
\label{eq:mu_rotation_secIV}
\end{equation}
with \(U_y=e^{-i\pi\sigma_y/4}\). Since \(U_y\) is independent of time, no additional gauge term is generated. The rotated Hamiltonian is
\begin{equation}
    \widetilde h_k(\mu)
    =
    2K(\mu+\cos k)\sigma_z
    +
    2K\sin k\,\sigma_x .
\label{eq:mu_rotated_hk_secIV}
\end{equation}
Thus
\begin{equation}
\begin{aligned}
    i\dot u_k&=a_k(\mu)u_k+b_kv_k,\\
    i\dot v_k&=b_ku_k-a_k(\mu)v_k,
\end{aligned}
\label{eq:mu_first_order_secIV}
\end{equation}
where
\begin{equation}
    a_k(\mu)=2K(\mu+\cos k),
    \quad
    b_k=2K\sin k .
\label{eq:mu_ab_def_secIV}
\end{equation}
Eliminating \(v_k\), we have
\begin{equation}
    \ddot u_k+
    \left[
    a_k^2(\mu)+b_k^2+i\dot a_k(\mu)
    \right]u_k=0 .
\label{eq:mu_second_time_secIV}
\end{equation}

For each linear segment of the quench, we write
\begin{equation}
    \frac{d\mu}{dt}
    =
    \frac{s_\mu}{\tau_s},
\label{eq:mu_linear_segment_secIV}
\end{equation}
where \(s_\mu=\pm1\) specifies the quench direction. The quench time is \(\tau_s=\tau_Q\) for the first critical dynamics and \(\tau_s=\tau_Q'=R\tau_Q\) for the second one. Using \(\mu\) as the evolution variable, Eq.~\eqref{eq:mu_second_time_secIV} becomes
\begin{equation}
    \frac{d^2u_k}{d\mu^2}
    +
    \left[
    \tau_s^2\varepsilon_k^2(\mu)
    +
    i\,s_\mu\,2K\tau_s
    \right]u_k=0,
\label{eq:mu_second_order_secIV}
\end{equation}
where
\begin{equation}
    \varepsilon_k(\mu)
    =
    2K\sqrt{(\mu+\cos k)^2+\sin^2 k}.
\label{eq:mu_epsilon_secIV}
\end{equation}
Equation~\eqref{eq:mu_second_order_secIV} is a Weber-type equation. This is the same differential-equation structure that appears in the Landau-Zener problem. Therefore, after two successive critical dynamics, the oscillatory part of the final occupation is controlled by a LZS phase~\cite{Shevchenko2010LZS}. The corresponding WKB construction is given in Sec.~S4 of the
Supplemental Material~\cite{SM}.

Within the WKB approximation, the oscillatory part of the bulk excitation density takes the form
\begin{equation}
    p_{\rm osc}^{\mathrm B}
    \sim
    \cos\left(
    \Delta\varphi_{\mathrm B}
    \right).
\label{eq:mu_bulk_osc_phase_relation_secIV}
\end{equation}
For protocol 3, the two critical dynamics occur at \(\mu=1\). Therefore, the interference phase is accumulated in the interval \(1<\mu<\mu_m\), namely the topologically trivial phase. The part before the first critical dynamics and the part after the second one change only the overall phase or the final phase offset, and do not enter the leading interference phase. Thus,
\begin{equation}
    \Delta\varphi_{\mathrm B}^{(3)}
    =
    (1+R)\tau_Q
    \int_{1}^{\mu_m}
    \Delta_{\mathrm B}(\mu)\,d\mu
    +
    \delta .
\label{eq:mu_bulk_phase_secIV}
\end{equation}
Here \(\delta\) denotes local connection phases and other terms that do not grow linearly with \(\tau_Q\).

The period is determined by requiring the interference phase to change by \(2\pi\),
\begin{equation}
    T_{\mathrm B}^{(3)}
    =
    \frac{2\pi}
    {
    \partial_{\tau_Q}\Delta\varphi_{\mathrm B}^{(3)}
    }.
\label{eq:mu_period_derivative_def_secIV}
\end{equation}
Using Eq.~\eqref{eq:mu_bulk_phase_secIV}, the period can be written directly in terms of the integrated energy gap as
\begin{equation}
    T_{\mathrm B}^{(3)}
    =
    \frac{2\pi}
    {
    (1+R)
    \int_{1}^{\mu_m}
    \Delta_{\mathrm B}(\mu)\,d\mu
    }.
\label{eq:mu_period_gap_relation_secIV}
\end{equation}
Equation~\eqref{eq:mu_period_gap_relation_secIV} shows that the interference period is fixed by the energy gap integrated in the topologically trivial region. The energy gap entering the WKB phase is
\begin{equation}
    \Delta_{\mathrm B}(\mu)
    = 4K\Lambda_{\pi}(\frac{\pi}{2},\mu)
    =
    4K(\mu-1).
\label{eq:mu_bulk_gap_secIV}
\end{equation}
Substituting Eq.~\eqref{eq:mu_bulk_gap_secIV} into Eq.~\eqref{eq:mu_period_gap_relation_secIV}, we obtain
\begin{equation}
    T_{\mathrm B}^{(3)}
    =
    \frac{\pi}
    {K(1+R)(\mu_m-1)^2}.
\label{eq:mu_TB_secIV}
\end{equation}
For \(K=1\), \(R=1\), and \(\mu_m=2.5\), this gives \(T_{\mathrm B}^{(3)}=2\pi/9\), 
which is the ICD period for bulk in Fig.~\ref{fig:ICDZM_patterns_main}(f).

The period-gap relation in Eq.~\eqref{eq:mu_period_derivative_def_secIV} also explains the ICDZM period in protocol 3. In contrast to protocols 1 and 2, the interference phase in Eq.~\eqref{eq:mu_bulk_phase_secIV} is accumulated in the topologically trivial phase. In this phase, the instantaneous \(j=1\) pair is not a zero-mode pair, and there is no zero mode to reduce the energy gap. Therefore, the interference phase is governed by the same energy gap as the bulk, i.e.,
\begin{equation}
    \Delta\varphi_{\mathrm Z}
    \simeq
    \Delta\varphi_{\mathrm B}^{(3)}
    +
    {\rm const.}
\label{eq:mu_Z_phase_secIV}
\end{equation}
Therefore,
\begin{equation}
    T_{\mathrm Z}
    \simeq
    T_{\mathrm B}.
\label{eq:mu_TZ_TB_secIV}
\end{equation}
This explains the same-period oscillation observed in Eq.~\eqref{eq:protocol3_period_relation_secIII}.

Combining the results of this subsection with those of Sec.~\ref{subsec:nontrivial_phase_secIV}, the ICDZM period is determined by the topological region in which the WKB phase is accumulated. When the phase is accumulated in the topologically nontrivial region, the zero-mode transfer channel can be governed by Eq.~\eqref{eq:theta_gap_half_secIV} and gives the possible period doubling \(T_{\mathrm Z}\simeq2T_{\mathrm B}\). When the phase is accumulated in the topologically trivial region, as in protocol 3, no such gap reduction occurs, and the zero-mode transfer probability follows the ordinary bulk ICD period.

\section{Boundary particle-number observable}
\label{sec:physical_observables}

In Sec.~\ref{sec:protocols_results}, the zero-mode transfer probability \(p^{\mathrm Z}\) is defined from the occupation of the final zero-mode pair. We now show that the same interference pattern can be obtained from the boundary particle number.

We define the rung-number operator
\begin{equation}
    \hat N_j
    =
    a_j^\dagger a_j+b_j^\dagger b_j .
\label{eq:phys_Nj_operator}
\end{equation}
For a quench process generated by one of the protocols, the initial and final expectation values of \(\hat N_j\) are defined as
\begin{equation}
    N_j^{\rm i}
    =
    \langle\psi(t_i)|\hat N_j|\psi(t_i)\rangle ,
    \quad
    N_j^{\rm f}
    =
    \langle\psi(t_f)|\hat N_j|\psi(t_f)\rangle .
\label{eq:phys_Nj_if_def}
\end{equation}
The deviation of the particle number on the \(j\)th rung is
\begin{equation}
    \Delta N_j
    =
    N_j^{\rm i}-N_j^{\rm f}.
\label{eq:phys_delta_Nj_def}
\end{equation}
For the selected initial ground state \(|\Psi_0\rangle\), the occupied zero mode is localized on the left boundary rung. We therefore define the left edge defect as
\begin{equation}
    d_{\rm left}
    =
    N_1^{\rm i}-N_1^{\rm f}
    =
    \Delta N_1 .
\label{eq:phys_dleft_def}
\end{equation}

At the initial point, the particle-number profile of the selected ground state is
\begin{equation}
    N_j^{\rm i}
    =
    \begin{cases}
    3/2, & j=1,\\
    1/2, & j=L,\\
    1, & j\neq1,L ,
    \end{cases}
\label{eq:phys_initial_density_profile}
\end{equation}
which carries the fractional boundary charges:
\begin{equation}
    Q_{\rm left}^{\rm i}
    =
    N_1^{\rm i}-\bar N
    =
    \frac12,
    \quad
    Q_{\rm right}^{\rm i}
    =
    N_L^{\rm i}-\bar N
    =
    -\frac12 ,
\label{eq:phys_initial_boundary_charge}
\end{equation}
where the bulk charge \(\bar N=1\) is obtained from the  particle-number conservation of PBC system. 

In the following, we use \(d_{\rm left}\) to characterize the change of the left boundary particle number during the quench dynamics.

\begin{figure}[t]
    \centering
    \includegraphics[width=1.0\linewidth]{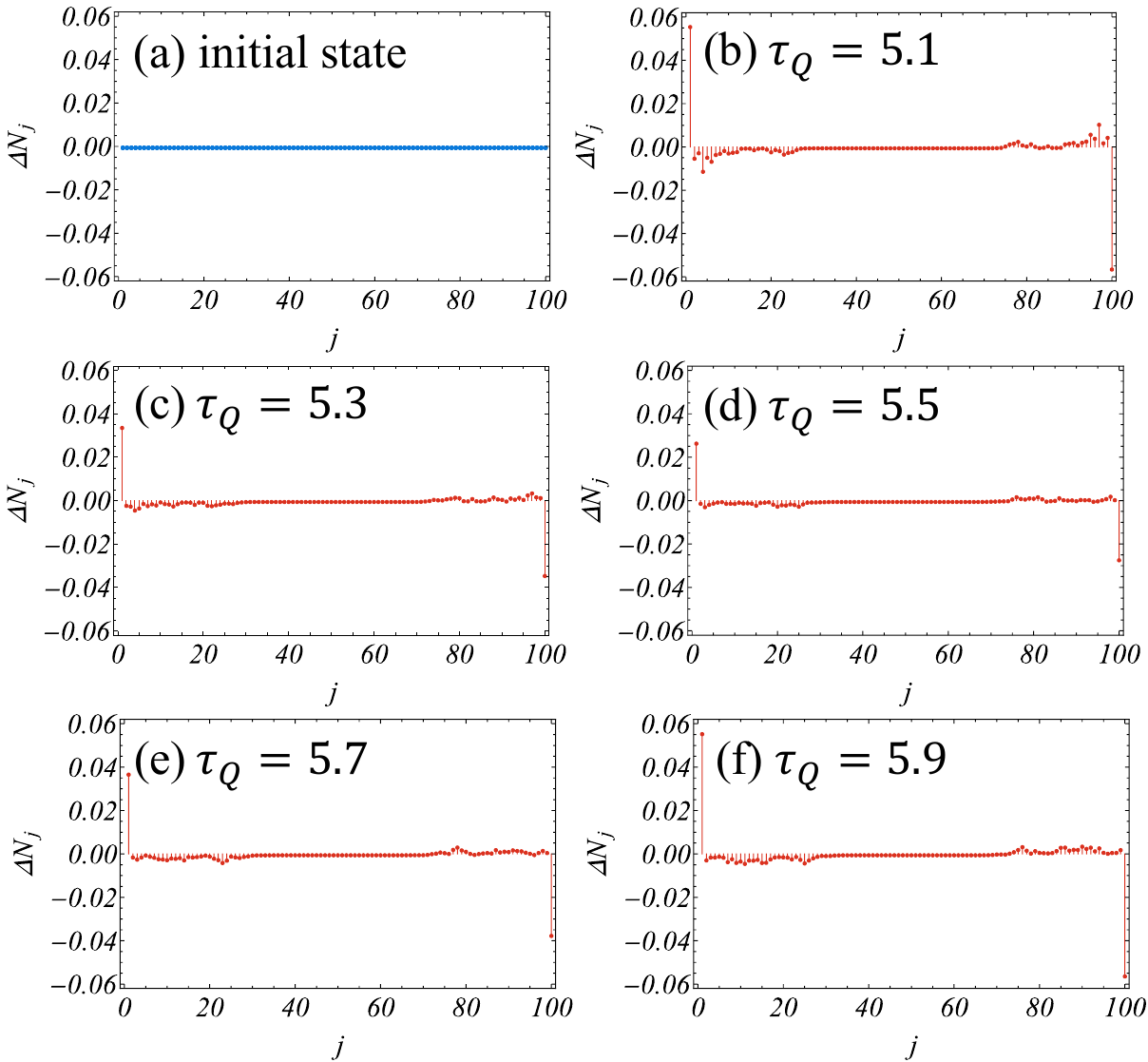}
    \caption{
    Deviations of the particle number produced in protocol 1.
    (a) Reference profile from the initial state, for which \(\Delta N_j=0\).
    (b)-(f) Deviations of the particle number
    \(\Delta N_j=N_j^{\rm i}-N_j^{\rm f}\)
    for \(\tau_Q=5.1,5.3,5.5,5.7,\) and \(5.9\), respectively.
    The left edge defect is the first-rung value
    \(d_{\rm left}=\Delta N_1\).
    }
    \label{fig:physical_observables_profile}
\end{figure}

Figure~\ref{fig:physical_observables_profile} shows the spatial profile of
\(\Delta N_j\) in protocol 1. The deviation of the particle number is concentrated near
the two boundaries, while the bulk rungs remain close to their initial value.
The first-rung value gives \(d_{\rm left}\), which is the boundary response associated with the initially occupied zero mode in \(|\Psi_0\rangle\). Together with the pair-occupatthe decrease of the particle number at the left boundary can be interpreted as the transfer probability of the selected zero mode. 

\begin{figure}[t]
    \centering
    \includegraphics[width=0.86\linewidth]{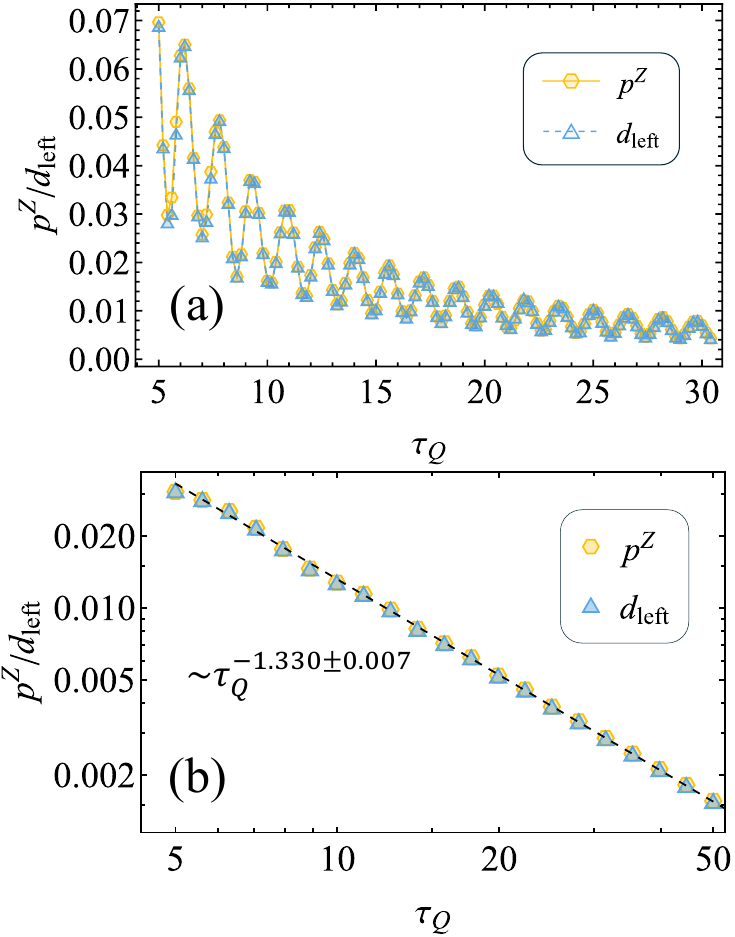}
    \caption{
    Comparison between the zero-mode transfer probability and the left edge
    defect.
    (a) The zero-mode transfer probability \(p^{\mathrm Z}\)
    and the left edge defect \(d_{\rm left}\) show the same oscillatory behavior as functions of \(\tau_Q\) after protocol 1.
    (b) One-way quench dynamics by varying \(\theta\) at fixed \(\mu=0\).
    The dashed line denotes the power-law fit
    \(d_{\rm left}\sim\tau_Q^{-\alpha_{\rm left}}\), with
    \(\alpha_{\rm left}=1.330\pm0.007\).
    }
    \label{fig:physical_observables_compare}
\end{figure}

Figure~\ref{fig:physical_observables_compare}(a) compares \(d_{\rm left}\) with \(p^{\mathrm Z}\) in protocol 1. The two quantities have the same oscillatory structure as functions of \(\tau_Q\). This means that the boundary particle-number change reproduces the ICDZM oscillation of the zero-mode transfer probability.

The same observable also works for the one-way quench dynamics by varying \(\theta\) at fixed \(\mu=0\). As shown in Fig.~\ref{fig:physical_observables_compare}(b), \(d_{\rm left}\) follows an algebraic decay in the fitting window,
\begin{equation}
    d_{\rm left}
    \sim
    \tau_Q^{-\alpha_{\rm left}} .
\label{eq:phys_dleft_oneway_scaling}
\end{equation}
The fit gives \(\alpha_{\rm left} =1.330\pm0.007\), which is close to the anomalous scaling reported for one-way Creutz-ladder quench dynamics~\cite{Bermudez2009PRL}. Therefore, the boundary particle-number change reproduces both the ICDZM oscillation in protocol 1 and the anomalous edge response in the one-way quench dynamics.

The quantities in Eq.~\eqref{eq:phys_Nj_if_def} are expectation values of local particle numbers. In site-resolved measurements of lattice systems, such expectation values are obtained by averaging repeated preparations \cite{Bakr2019,Bloch2022}. In one realization, the first-rung occupation is integer valued,
\begin{equation}
    N_{1,\mathrm{exp}}
    =
    0,\ 1,\ \mathrm{or}\ 2 ,
\label{eq:phys_single_shot_N1}
\end{equation}
because the first rung contains two fermionic orbitals \(a_1\) and \(b_1\). The expectation values are obtained as
\begin{equation}
    N_1^{\rm i}
    =
    \overline{N_{1,\mathrm{exp}}^{\rm i}},
    \qquad
    N_1^{\rm f}
    =
    \overline{N_{1,\mathrm{exp}}^{\rm f}},
\label{eq:phys_exp_average}
\end{equation}
where the overline denotes the average over repeated preparations. Therefore,
\begin{equation}
    d_{\rm left}
    =
    \overline{N_{1,\mathrm{exp}}^{\rm i}}
    -
    \overline{N_{1,\mathrm{exp}}^{\rm f}} .
\label{eq:phys_dleft_exp_average}
\end{equation}
The initial average is obtained by preparing \(|\Psi_0\rangle\) and measuring the first-rung occupation. The final average is obtained by preparing the same initial state, applying the quench, and measuring the first-rung occupation at \(t_f\).

If the other half-filled ground state \(|\Psi_0'\rangle\) is chosen, the initially occupied zero mode is localized on the right boundary. The corresponding edge defect is then
\begin{equation}
    d_{\rm right}
    =
    N_L^{\rm i}-N_L^{\rm f}.
\label{eq:phys_dright_def}
\end{equation}
The right-boundary result is discussed in Appendix~\ref{app:physical_observables_right}.

\section{Conclusion}
\label{sec:conclusion}

We have investigated the interference of critical dynamics associated with zero modes in the generalized Creutz ladders. By employing three closed quench paths, we showed that the ICDZM pattern obtained from the zero-mode transfer probability retains information about the topological region visited between two critical dynamics. When the path passes through the topologically nontrivial phase with the opposite winding number, the ICDZM pattern exhibits period doubling compared with the bulk ICD period. When the path crosses the same critical point twice, the ICDZM pattern is strongly suppressed, although the bulk excitation density remains oscillatory. When the path enters the topologically trivial region, the final zero-mode pair becomes approximately balanced, and the ICDZM period follows the bulk ICD period. Within the framework of WKB analysis, these different ICDZM patterns are ascribed as the interference phase accumulated during the quench and by the energy gap entering this phase. Thus, the oscillation period of the ICDZM pattern reveals whether the phase is accumulated in a topologically nontrivial or trivial region. 

We further showed that the zero-mode transfer prob- ability can be diagnosed through an edge defect defined from the boundary particle number, which also connects to the initial fractional boundary charge. These results identify ICDZM and the corresponding edge defect as probes for critical dynamics associated with topological zero modes.

\section*{ACKNOWLEDGMENTS}
We thank  Yan He and Wenxing Nie for insightful discussions. This work is supported by Natural Science Foundation of Sichuan Province (2026NSFSC0772).

\appendix

\section{Right-edge local observable for the second selected ground state}
\label{app:physical_observables_right}

In Sec.~\ref{sec:physical_observables}, we used the selected ground state \(|\Psi_0\rangle\), for which the occupied zero mode is localized at the left boundary at the common Creutz endpoint. If the other member of the zero-mode doublet, \(|\Psi_0'\rangle\), is selected instead, the same construction applies at the opposite boundary. The zero-mode transfer probability is defined in the same way as in Eq.~\eqref{eq:pZ_main_def}, but with the corresponding final zero mode selected by \(|\Psi_0'\rangle\).

The local density on the right boundary rung is
\begin{equation}
    N_{\rm end}(t)
    =
    \left\langle
    a_L^\dagger a_L+b_L^\dagger b_L
    \right\rangle_t .
    \label{eq:app_phys_Nend_def}
\end{equation}
We then define the right-edge density defect as
\begin{equation}
    d_{\rm right}(t_f)
    =
    N_{\rm end}(t_i)-N_{\rm end}(t_f).
    \label{eq:app_phys_dedge_right_def}
\end{equation}
This quantity is the right-edge counterpart of
\(d_{\rm edge}^{(\mathrm{l})}\) defined in Eq.~\eqref{eq:phys_dleft_def}.

As in the left-edge case, \(N_{\rm end}(t)\) is an expectation value of a local number operator. Its experimental interpretation is therefore obtained by averaging over repeated preparations and measurements performed under the same conditions.
\begin{equation}
    N_{\rm end}(t)
    =
    \overline{N_{\rm end}^{\rm exp}(t)} ,
    \label{eq:app_phys_Nend_exp_average}
\end{equation}
where \(N_{\rm end}^{\rm exp}=0,1,\) or \(2\) in each single realization. Accordingly,
\begin{equation}
    d_{\rm right}
    =
    \overline{N_{\rm end}^{\rm exp}(t_i)}
    -
    \overline{N_{\rm end}^{\rm exp}(t_f)} .
    \label{eq:app_phys_dedge_right_exp_average}
\end{equation}
The initial average can be calibrated by preparing \(|\Psi_0'\rangle\) and measuring the right-boundary rung without applying the quench. The final average is obtained from independent repetitions in which the same initial state is prepared, evolved through the quench, and measured at the final time.


The density defect \(d_{\rm right}\) carries the same oscillatory behavior due to ICDZM as the zero-mode transfer probability obtained from the selected state \(|\Psi_0'\rangle\). Thus the boundary used in the local particle-number measurement is tied to which zero mode is occupied initially: \(d_{\rm left}\) for \(|\Psi_0\rangle\), and \(d_{\rm right}\) for \(|\Psi_0'\rangle\).

\bibliographystyle{apsrev4-2}
\bibliography{citation}

\clearpage
\onecolumngrid

\begingroup

\setcounter{section}{0}
\setcounter{subsection}{0}
\setcounter{equation}{0}
\setcounter{figure}{0}
\setcounter{table}{0}

\renewcommand{\thesection}{S\arabic{section}}
\renewcommand{\thesubsection}{\thesection.\arabic{subsection}}
\renewcommand{\theequation}{S\arabic{equation}}
\renewcommand{\thefigure}{S\arabic{figure}}
\renewcommand{\thetable}{S\arabic{table}}

\newcommand{\SMsection}[1]{%
    \refstepcounter{section}%
    \setcounter{subsection}{0}%
    \section*{\thesection. #1}%
}

\newcommand{\SMsubsection}[1]{%
    \refstepcounter{subsection}%
    \subsection*{\thesubsection. #1}%
}

\begin{center}
{\large\bf Supplemental Material for\\
``Interference of critical dynamics associated with zero modes''\par}
\vspace{0.7em}
\end{center}

\vspace{1em}

\noindent
This Supplemental Material provides the details supporting the main text. In Sec.~S1 we describe the open-boundary numerical construction and the instantaneous-mode decomposition. In Sec.~S2 we prove the pair-occupation relation used to define the zero-mode depletion. In Secs.~S3--S5 we derive the effective differential equations and WKB phases for the three protocols.

\vspace{1em}

\SMsection{Open-boundary numerical evaluation and instantaneous-mode distributions}
\label{sec:supp_obc_numerics}

This section gives the technical details used to evaluate the open-boundary observables in the main text. The formulas below are not needed for the physical discussion in the main text, but they specify how the many-body quantities are reduced to single-particle evolution for the particle-number-conserving quadratic Hamiltonian.

\SMsubsection{Single-particle evolution in the open chain}
\label{subsec:supp_obc_single_particle}

For a fixed value of the quench parameter \(\lambda\), with \(\lambda=\mu\) or \(\theta\), the OBC Hamiltonian is diagonalized as
\begin{equation}
    H(\lambda)
    =
    \sum_{j=1}^{L}
    \left[
    E_{j,+}(\lambda)\eta_{j,+}^{\dagger}(\lambda)\eta_{j,+}(\lambda)
    -
    E_{j,-}(\lambda)\eta_{j,-}^{\dagger}(\lambda)\eta_{j,-}(\lambda)
    \right],
\label{eq:supp_obc_diag}
\end{equation}
where \(E_{j,+}\ge0\), \(E_{j,-}\ge0\), and the single-particle modes are ordered by increasing absolute energy. The modes are ordered by increasing absolute energy, \(0\le E_{1,\pm}\le E_{2,\pm}\le\cdots\).The corresponding quasiparticle operator is written as
\begin{equation}
    \eta_{\beta}(\lambda)
    =
    \sum_{n=1}^{L}
    \left[
    u_{\beta,n}^{*}(\lambda)a_n
    +
    v_{\beta,n}^{*}(\lambda)b_n
    \right],
    \qquad
    \beta=(j,\pm).
\label{eq:supp_obc_eta_beta}
\end{equation}
The half-filled initial state used in the main text is
\begin{equation}
    |\Psi_0\rangle
    =
    \prod_{j=1}^{L}
    \eta_{j,-}^{\dagger}(\lambda_i)|0\rangle,
\label{eq:supp_obc_initial_state}
\end{equation}
so that the initially occupied set is
\begin{equation}
    {\rm occ}=\{(j,-)\,|\,j=1,\cdots,L\}.
\label{eq:supp_obc_occ_set}
\end{equation}

For an initially occupied single-particle mode \(\beta\in{\rm occ}\), we define the evolved creation operator by
\begin{equation}
    \widetilde{\eta}_{\beta}^{\dagger}(t)
    =
    U(t,t_i)\eta_{\beta}^{\dagger}(\lambda_i)U^{\dagger}(t,t_i),
\label{eq:supp_evolved_eta_def}
\end{equation}
where \( U(t,t_i)=\mathcal{T}\exp\left[-i\int_{t_i}^{t}H(\tau)\,d\tau\right]\) is the time-evolution operator. Since the Hamiltonian is quadratic and conserves particle number, \(\widetilde{\eta}_{\beta}^{\dagger}(t)\) remains in the one-particle space and can be expanded as
\begin{equation}
    \widetilde{\eta}_{\beta}^{\dagger}(t)
    =
    \sum_{n=1}^{L}
    \left[
    \widetilde u_{\beta,n}(t)a_n^{\dagger}
    +
    \widetilde v_{\beta,n}(t)b_n^{\dagger}
    \right].
\label{eq:supp_evolved_eta_uv}
\end{equation}
The initial condition is
\begin{equation}
    \widetilde u_{\beta,n}(t_i)=u_{\beta,n}(\lambda_i),
    \qquad
    \widetilde v_{\beta,n}(t_i)=v_{\beta,n}(\lambda_i).
\label{eq:supp_evolved_eta_initial}
\end{equation}
The amplitudes are obtained from
\begin{equation}
    i\frac{d}{dt}\widetilde{\eta}_{\beta}^{\dagger}(t)
    =
    \left[H(t),\widetilde{\eta}_{\beta}^{\dagger}(t)\right].
\label{eq:supp_heisenberg_eta}
\end{equation}
For the parameter plane \(J_X=J_D=K\), \(\phi=0\), and \(\mu=J_Y/(2K)\), this gives
\begin{equation}
\begin{aligned}
\frac{1}{iK}\frac{d}{dt}\widetilde u_{\beta,n}(t)
={}&
 e^{i\theta(t)}\widetilde u_{\beta,n-1}(t)
+
 e^{-i\theta(t)}\widetilde u_{\beta,n+1}(t)
\\
&+2\mu(t)\widetilde v_{\beta,n}(t)
+
\widetilde v_{\beta,n-1}(t)
+
\widetilde v_{\beta,n+1}(t),
\\[1.0ex]
\frac{1}{iK}\frac{d}{dt}\widetilde v_{\beta,n}(t)
={}&
 e^{-i\theta(t)}\widetilde v_{\beta,n-1}(t)
+
 e^{i\theta(t)}\widetilde v_{\beta,n+1}(t)
\\
&+2\mu(t)\widetilde u_{\beta,n}(t)
+
\widetilde u_{\beta,n-1}(t)
+
\widetilde u_{\beta,n+1}(t).
\end{aligned}
\label{eq:supp_real_space_eom}
\end{equation}
Open boundaries are imposed by setting all amplitudes outside the chain to zero,
\begin{equation}
    \widetilde u_{\beta,0}=\widetilde u_{\beta,L+1}
    =
    \widetilde v_{\beta,0}=\widetilde v_{\beta,L+1}=0 .
\label{eq:supp_obc_boundary_condition}
\end{equation}
Equation~\eqref{eq:supp_real_space_eom} is the equation used in the numerical time evolution.

\SMsubsection{Evaluation of \texorpdfstring{\(p^{ Z}\)}{pz} and \texorpdfstring{\(p^{\mathrm B}\)}{pb}}
\label{subsec:supp_obc_PZ_PB}

It is useful to introduce the single-particle transfer amplitude from an initial mode \(\beta\) to an instantaneous mode \(\alpha\) at time \(t\):
\begin{equation}
\begin{aligned}
    A_{\alpha\leftarrow\beta}(t)
    &=
    \langle0|\eta_{\alpha}(\lambda(t))
    U(t,t_i)\eta_{\beta}^{\dagger}(\lambda_i)|0\rangle
    \\
    &=
    \sum_{n=1}^{L}
    \left[
    u_{\alpha,n}^{*}(\lambda(t))\widetilde u_{\beta,n}(t)
    +
    v_{\alpha,n}^{*}(\lambda(t))\widetilde v_{\beta,n}(t)
    \right].
\end{aligned}
\label{eq:supp_A_alpha_beta_t}
\end{equation}
Here \(\alpha=(j,\pm)\) labels the instantaneous OBC eigenmode at \(\lambda(t)\), while \(\beta\) labels the initial eigenmode at \(\lambda_i\). The normalization of the one-particle evolution gives
\begin{equation}
    \sum_{\alpha}|A_{\alpha\leftarrow\beta}(t)|^2=1
\label{eq:supp_A_normalization}
\end{equation}
for each initial mode \(\beta\).

At the final Creutz endpoint used in the main text, the selected occupied zero mode is
\begin{equation}
    \eta_{1,-}^{\dagger}(\lambda_f)
    =
    \frac{1}{\sqrt2}(a_1^{\dagger}-i b_1^{\dagger}).
\label{eq:supp_final_zero_mode}
\end{equation}
Therefore,
\begin{equation}
    A_{(1,-)\leftarrow\beta}(t_f)
    =
    \frac{1}{\sqrt2}
    \left[
    \widetilde u_{\beta,1}(t_f)
    +
    i\widetilde v_{\beta,1}(t_f)
    \right].
\label{eq:supp_A_zf_beta_boundary}
\end{equation}
The final occupation of the selected zero mode is
\begin{equation}
    n_-(t_f)
    =
    \sum_{\beta\in{\rm occ}}
    |A_{(1,-)\leftarrow\beta}(t_f)|^2,
\label{eq:supp_nminus_A}
\end{equation}
and hence
\begin{equation}
    p^{Z}(t_f)
    =
    1-
    \sum_{\beta\in{\rm occ}}
    |A_{(1,-)\leftarrow\beta}(t_f)|^2 .
\label{eq:supp_PZ_A}
\end{equation}
Equivalently, using Eq.~\eqref{eq:supp_A_zf_beta_boundary},
\begin{equation}
    p^{Z}(t_f)
    =
    1-
    \frac12
    \sum_{\beta\in{\rm occ}}
    \left|
    \widetilde u_{\beta,1}(t_f)
    +
    i\widetilde v_{\beta,1}(t_f)
    \right|^2 .
\label{eq:supp_PZ_boundary_uv}
\end{equation}
The bulk excitation density is evaluated from the final positive-branch occupations,
\begin{equation}
    p^{\mathrm B}(t_f)
    =
    \frac{1}{L}
    \sum_{j=1}^{L}
    \sum_{\beta\in{\rm occ}}
    |A_{(j,+)\leftarrow\beta}(t_f)|^2 .
\label{eq:supp_PB_A}
\end{equation}
Equations~\eqref{eq:supp_PZ_A} and \eqref{eq:supp_PB_A} are the numerical formulas corresponding to the many-body expectation values in the main text.

\SMsubsection{Extraction of the oscillatory contribution}
\label{subsec:supp_obc_nosc}

For the closed quench protocols, the evolution contains two consecutive critical dynamics. To separate the final zero-mode occupation into intensity and oscillatory contributions, we insert a complete instantaneous single-particle basis at an intermediate time \(t_M\). In the protocols considered in the main text, \(t_M=0\) is chosen at the turning point of the quench path. We write
\begin{equation}
    U(t_f,t_i)=U(t_f,t_M)U(t_M,t_i)\equiv U_2U_1,
\label{eq:supp_two_stage_U}
\end{equation}
and define the two-stage transfer amplitudes as
\begin{equation}
    A_{\gamma\leftarrow\beta}^{(1)}
    =
    \langle0|\eta_{\gamma}(\lambda_M)
    U(t_M,t_i)\eta_{\beta}^{\dagger}(\lambda_i)|0\rangle,
\label{eq:supp_A1_def}
\end{equation}
\begin{equation}
A_{(1,-)\leftarrow\gamma}^{(2)}
=
\langle0|\eta_{1,-}(\lambda_f)
U(t_f,t_M)\eta_{\gamma}^{\dagger}(\lambda_M)|0\rangle .
\label{eq:supp_A2_def}
\end{equation}
where \(\lambda_M=\lambda(t_M)\). Inserting the complete basis at \(t_M\) gives
\begin{equation}
    A_{(1,-)\leftarrow\beta}(t_f)
    =
    \sum_{\gamma}
    A_{(1,-)\leftarrow\gamma}^{(2)}A_{\gamma\leftarrow\beta}^{(1)},
\label{eq:supp_A_two_stage}
\end{equation}
where \(\gamma\) runs over all instantaneous OBC eigenmodes at \(\lambda_M\). Substituting Eq.~\eqref{eq:supp_A_two_stage} into Eq.~\eqref{eq:supp_nminus_A}, one obtains
\begin{equation}
\begin{split}
    n_-(t_f)
    &=
    \sum_{\beta\in{\rm occ}}
    \left|
    \sum_{\gamma}A_{(1,-)\leftarrow\gamma}^{(2)}A_{\gamma\leftarrow\beta}^{(1)}
    \right|^2
    \\
    &=
    n_0(t_f)+n_{\rm osc}(t_f).
\end{split}
\label{eq:supp_nminus_decomp}
\end{equation}
The intensity contribution is
\begin{equation}
    n_0(t_f)
    =
    \sum_{\beta\in{\rm occ}}
    \sum_{\gamma}
    \left|
    A_{(1,-)\leftarrow\gamma}^{(2)}A_{\gamma\leftarrow\beta}^{(1)}
    \right|^2 .
\label{eq:supp_n0}
\end{equation}
The oscillatory contribution is the coherent cross-term part,
\begin{equation}
\begin{split}
    n_{\rm osc}(t_f)
    =
    2\,\operatorname{Re}
    \sum_{\beta\in{\rm occ}}
    \sum_{\gamma<\gamma'}
    &A_{(1,-)\leftarrow\gamma}^{(2)}A_{\gamma\leftarrow\beta}^{(1)}
    \\
    &\times
    A_{(1,-)\leftarrow\gamma'}^{(2)*}A_{\gamma'\leftarrow\beta}^{(1)*} .
\end{split}
\label{eq:supp_nosc}
\end{equation}
Equivalently, one may write the same term as a sum over \(\gamma\neq\gamma'\). The form in Eq.~\eqref{eq:supp_nosc} makes the reality of \(n_{\rm osc}\) explicit. This is the quantity plotted in the insets of the zero-mode-depletion figures in the main text.

\SMsection{Mode Pairing during the quench}
\label{sec:supp_pair_occupation_sum_rule}
For the symmetry-preserving quench paths considered below, each instantaneous OBC pair \((j,+),(j,-)\) satisfies a pair-occupation sum rule in the half-filled state. We define the operator
\begin{equation}
    \hat{\rho}_j(t)
    =
    \hat n_{j,+}(t)+\hat n_{j,-}(t),
\end{equation}
where
\begin{equation}
    \hat n_{j,\pm}(t)
    =
    \eta_{j,\pm}^{\dagger}(t)\eta_{j,\pm}(t).
\end{equation}
Our aim is to prove
\begin{equation}
    \langle \hat{\rho}_j(t)\rangle
    =
    \langle \Psi_0(t)|
    \eta_{j,+}^{\dagger}(t)\eta_{j,+}(t)
    +
    \eta_{j,-}^{\dagger}(t)\eta_{j,-}(t)
    |\Psi_0(t)\rangle
    =
    1 .
\label{eq:supp_pair_sum_target}
\end{equation}
For \(j=1\), this gives the zero-mode-pair relation
\begin{equation}
    N_{\mathrm Z}(t)=n_{1,+}(t)+n_{1,-}(t)= 1.
\end{equation}

\SMsubsection{Symmetries of the two quench paths}

We first specify the symmetry operations used below. These operations are
written directly in the \(a_l,b_l\) basis.

\SMsubsection{\(\theta\)-quench}

For protocols 1 and 2, the quench is performed along
\begin{equation}
    \theta(t)\sim\frac{t}{\tau_Q}.
\end{equation}
with \(J_Y=0\). This is the rungless Creutz-ladder line. Following the hidden-symmetry structure discussed for Creutz-type ladders in Ref.~\cite{Zurita2021Quantum}, we introduce
\begin{equation}
    s_l=(-1)^{l-1}.
\end{equation}
The hidden chiral operation \(X_S\) acts as
\begin{equation}
    X_S a_l^\dagger X_S^{-1}
    =
    s_l a_l^\dagger,
    \qquad
    X_S b_l^\dagger X_S^{-1}
    =
    s_l b_l^\dagger .
\label{eq:supp_XS_action}
\end{equation}
Equivalently, in matrix notation,
\begin{equation}
    X_S = \mathrm{diag}_{2L}(1,1,-1,-1,1,1,-1,-1,\cdots).
\end{equation}
in basis \((a_1,b_1,\cdots,a_L,b_L)\), and we have
\begin{equation}
    X_S H_{\theta}(t)X_S^{-1}
    =
    -H_{\theta}(t),
\label{eq:supp_XS_chiral_theta}
\end{equation}
Thus, if the quasi-particle operator for a positive energy level can be written as:
\begin{equation}
    \eta_{j,+}^{\dagger}(t)
    =
    \sum_{l=1}^{L}
    \left[
    u_{j,l}(t)a_l^\dagger
    +
    v_{j,l}(t)b_l^\dagger
    \right],
\end{equation}
then the hidden-chiral partner should be chosen as
\begin{equation}
    \eta_{j,-}^{\dagger}(t)
    =
    \sum_{l=1}^{L}
    s_l
    \left[
    u_{j,l}(t)a_l^\dagger
    +
    v_{j,l}(t)b_l^\dagger
    \right].
\label{eq:supp_theta_XS_pair}
\end{equation}
This is the coefficient relation protected by \(X_S\).

For the critical dynamics, the time-reversed symmetry is broken, so we also need the hidden particle-hole operation
\begin{equation}
    C_{SC}=R_{SC}K ,
\end{equation}
where \(K\) complex-conjugates coefficients and
\begin{equation}
    R_{SC}a_l^\dagger R_{SC}^{-1}
    =
    s_l b_l^\dagger,
    \qquad
    R_{SC}b_l^\dagger R_{SC}^{-1}
    =
    s_l a_l^\dagger .
\label{eq:supp_RSC_action}
\end{equation}
Equivalently,
\begin{equation}
    R_{SC}
    =
    \mathrm{diag}_{L}
    (\sigma_x,-\sigma_x,\sigma_x,-\sigma_x,\cdots).
\end{equation}
This operation is also the hidden particle-hole operation listed for the Creutz-type ladder in Ref.~\cite{Zurita2021Quantum}. Acting on the Hamiltonian, one obtains
\begin{equation}
    R_{SC}H_{\theta}^{*}(t)R_{SC}^{-1}
    =
    -H_{\theta}(t),
\label{eq:supp_RSC_ph_theta}
\end{equation}
Indeed, under \(C_{SC}\), for example,
\begin{equation}
    e^{i\theta}a_{l+1}^{\dagger}a_l
    \longrightarrow
    e^{-i\theta}s_{l+1}s_l b_{l+1}^{\dagger}b_l
    =
    -e^{-i\theta}b_{l+1}^{\dagger}b_l ,
\end{equation}
which is the negative of the corresponding \(b\)-leg hopping. The diagonal hoppings transform in the same way, while no rung term is present for \(J_Y=0\).

Equation~\eqref{eq:supp_RSC_ph_theta} also constrains the time-evolution operator. Let \(U(t,t_i)\) be the single-particle evolution operator,
\[
i\partial_t U(t,t_i)=h_\theta(t)U(t,t_i),
\]
where \(U(t,t_i)=\exp[-\mathcal{T}\int_{t_i}^{t}H(\tau)d\tau]\) is time-evolution operator. Taking the complex conjugate gives
\[
i\partial_t U^*(t,t_i)=-h_\theta^*(t)U^*(t,t_i).
\]
Define
\[
\widetilde U(t,t_i)=R_{SC}U^*(t,t_i)R_{SC}^{-1}.
\]
Since \(R_{SC}\) is time independent,
\[
i\partial_t\widetilde U
=
-\left(R_{SC}h_\theta^*R_{SC}^{-1}\right)\widetilde U.
\]
Using
\[
R_{SC}h_\theta^*(t)R_{SC}^{-1}=-h_\theta(t),
\]
we obtain
\[
i\partial_t\widetilde U=h_\theta(t)\widetilde U.
\]
Moreover, \(\widetilde U(t_i,t_i)=I\).

For nonzero-energy modes, both \(X_S\) and \(C_{SC}\) map an eigenmode to the opposite-energy partner. Their results can differ only by a phase. For the zero-mode doublet, the individual basis is not unique, but the total operator \(\hat{\rho}_Z=\eta_{1,+}^\dagger\eta_{1,+}+\eta_{1,-}^\dagger\eta_{1,-}\) is invariant under any unitary rotation inside the two-dimensional zero-mode subspace.

\SMsubsection{\(J_Y\)-quench}

For protocol 3, we set
\begin{equation}
    J_Y(t)\sim\frac{t}{\tau_Q}.
\end{equation}
with \(\theta=\pi/2\). The Hamiltonian has the ordinary internal chiral symmetry
\begin{equation}
    \Gamma_Y=I_L\otimes\sigma_y ,
\end{equation}
or, in the \(a_l,b_l\) basis,
\begin{equation}
    \Gamma_Y a_l^\dagger \Gamma_Y^{-1}
    =
    i b_l^\dagger,
    \qquad
    \Gamma_Y b_l^\dagger \Gamma_Y^{-1}
    =
    -i a_l^\dagger .
\end{equation}
Equivalently, the corresponding bilinear one-particle operator is
\begin{equation}
    \widehat{\Gamma}_Y
    =
    \sum_{l=1}^{L}
    \left(
    -i a_l^\dagger b_l
    +
    i b_l^\dagger a_l
    \right).
\end{equation}
For \(\theta=\pi/2\) and \(\phi=0\), one has
\begin{equation}
    \Gamma_Y H_Y(t)\Gamma_Y^{-1}
    =
    -H_Y(t).
\label{eq:supp_GammaY_chiral}
\end{equation}
This is the usual \(\sigma_y\)-type chiral symmetry of the Creutz ladder
\cite{Roy2023PRB,Chiu2016RMP}.

For the critical dynamics, it is useful to introduce antiunitary operation:
\begin{equation}
    C_Y=R_YK,
    \qquad
    R_Y=I_L\otimes\sigma_z ,
\end{equation}
where \(K\) denotes complex conjugation in the \(a_l,b_l\) basis. Explicitly,
\begin{equation}
    R_Y a_l^\dagger R_Y^{-1}
    =
    a_l^\dagger,
    \qquad
    R_Y b_l^\dagger R_Y^{-1}
    =
    -b_l^\dagger .
\label{eq:supp_RY_action}
\end{equation}
A direct term-by-term check gives
\begin{equation}
    R_Y H_Y^*(t)R_Y^{-1}
    =
    -H_Y(t).
\label{eq:supp_RY_ph}
\end{equation}
Indeed, complex conjugation reverses the phases of the leg hoppings at \(\theta=\pi/2\), while \(R_Y\) changes the sign of all terms containing one \(b_l\) operator. Therefore the leg hoppings, the diagonal hoppings, and the rung \(J_Y\) term are all mapped to the negative of themselves.

Equation~\eqref{eq:supp_RY_ph} implies
\begin{equation}
    R_YU^*(t,t_i)R_Y^{-1}
    =
    U(t,t_i).
\label{eq:supp_RY_U_relation}
\end{equation}

\SMsubsection{Occupation in the half-filled state}

We now evaluate the observable \(\hat{\rho}_j(t)\) in the physical state.
Define the evolved initial single-particle operator
\begin{equation}
    \widetilde{\eta}_{m,\tau}^{\dagger}(t)
    =
    U(t,t_i)\eta_{m,\tau}^{\dagger}(t_i)U^\dagger(t,t_i),
\end{equation}
with \(\tau=\pm\). Since the Hamiltonian is quadratic and particle-number conserving,
\(\widetilde{\eta}_{m,\tau}^{\dagger}(t)\) remains a linear combination of
\(a_l^\dagger\) and \(b_l^\dagger\). We use the single-particle transfer amplitude in Eq.~\eqref{eq:supp_A_alpha_beta_t}
\begin{equation}
      A_{j\sigma\leftarrow m\tau}(t)
    =
    \langle0|
    \eta_{j,\sigma}(t)
    \widetilde{\eta}_{m,\tau}^{\dagger}(t)
    |0\rangle .
\label{eq:supp_transfer_amp_def}
\end{equation}
This is the amplitude that an evolved initial mode \((m,\tau)\) is detected in the instantaneous mode \((j,\sigma)\).

Because the physical initial state fills only the initial \((-)\) modes,
\begin{equation}
    |\Psi_0\rangle
    =
    \prod_{m=1}^{L}
    \eta_{m,-}^{\dagger}(t_i)|0\rangle ,
\end{equation}
the occupation of the instantaneous mode \((j,\sigma)\) is
\begin{equation}
    n_{j,\sigma}(t)
    =
    \langle\Psi_0(t)|
    \eta_{j,\sigma}^{\dagger}(t)\eta_{j,\sigma}(t)
    |\Psi_0(t)\rangle
    =
    \sum_{m=1}^{L}
    \left|
      A_{j\sigma\leftarrow m-}(t)
    \right|^2 .
\label{eq:supp_occ_transfer}
\end{equation}
Therefore,
\begin{equation}
    \langle\hat\rho_j(t)\rangle
    =
    \sum_m
    \left|
      A_{j+\leftarrow m-}(t)
    \right|^2
    +
    \sum_m
    \left|
      A_{j-\leftarrow m-}(t)
    \right|^2 .
\label{eq:supp_rho_transfer}
\end{equation}

We next prove a universal completeness identity. Write the measured mode as
\begin{equation}
    \eta_{j,-}^{\dagger}(t)
    =
    \sum_{l=1}^{L}
    \left[
    u_{j,l}(t)a_l^\dagger
    +
    v_{j,l}(t)b_l^\dagger
    \right],
\label{eq:supp_measured_eta_minus}
\end{equation}
and write the complete evolved initial basis as
\begin{equation}
    \widetilde{\eta}_{m,\tau}^{\dagger}(t)
    =
    \sum_{l=1}^{L}
    \left[
    \widetilde u_{m\tau,l}(t)a_l^\dagger
    +
    \widetilde v_{m\tau,l}(t)b_l^\dagger
    \right],
\label{eq:supp_tilde_eta_uv}
\end{equation}
Then
\begin{equation}
      A_{j-\leftarrow m\tau}(t)
    =
    \sum_{l=1}^{L}
    \left[
    u_{j,l}^{*}(t)\widetilde u_{m\tau,l}(t)
    +
    v_{j,l}^{*}(t)\widetilde v_{m\tau,l}(t)
    \right].
\label{eq:supp_amp_uv}
\end{equation}
The \(2L\) evolved modes \(\{\widetilde{\eta}_{m,\tau}^{\dagger}(t)\}\) form a complete orthonormal basis of the one-particle space. Hence their coefficients obey
\begin{align}
    \sum_{m,\tau}
    \widetilde u_{m\tau,l}
    \widetilde u_{m\tau,r}^{*}
    &=
    \delta_{lr},
\\
    \sum_{m,\tau}
    \widetilde v_{m\tau,l}
    \widetilde v_{m\tau,r}^{*}
    &=
    \delta_{lr},
\\
    \sum_{m,\tau}
    \widetilde u_{m\tau,l}
    \widetilde v_{m\tau,r}^{*}
    &=
    0,
\\
    \sum_{m,\tau}
    \widetilde v_{m\tau,l}
    \widetilde u_{m\tau,r}^{*}
    &=
    0.
\end{align}
Using these relations, we find
\begin{align}
&\sum_{m,\tau}
\left|
  A_{j-\leftarrow m\tau}(t)
\right|^2
\nonumber\\
=&
\sum_{l,r}
u_{j,l}^{*}u_{j,r}
\sum_{m,\tau}
\widetilde u_{m\tau,l}\widetilde u_{m\tau,r}^{*}
+
\sum_{l,r}
v_{j,l}^{*}v_{j,r}
\sum_{m,\tau}
\widetilde v_{m\tau,l}\widetilde v_{m\tau,r}^{*}
\nonumber\\
&+
\sum_{l,r}
u_{j,l}^{*}v_{j,r}
\sum_{m,\tau}
\widetilde u_{m\tau,l}\widetilde v_{m\tau,r}^{*}
+
\sum_{l,r}
v_{j,l}^{*}u_{j,r}
\sum_{m,\tau}
\widetilde v_{m\tau,l}\widetilde u_{m\tau,r}^{*}
\nonumber\\
=&
\sum_l
\left(
|u_{j,l}(t)|^2
+
|v_{j,l}(t)|^2
\right)
=
1 .
\label{eq:supp_completeness_sum}
\end{align}
Therefore,
\begin{equation}
    \sum_m
    \left|
      A_{j-\leftarrow m-}(t)
    \right|^2
    +
    \sum_m
    \left|
      A_{j-\leftarrow m+}(t)
    \right|^2
    =
    1 ,
\label{eq:supp_missing_contribution}
\end{equation}
or
\begin{equation}
    \sum_{m,\sigma}
    \left|
    \bra{0}\eta_{j,-}(t)
    U(t,t_i)
    \eta_{m,\sigma}^\dagger(t_i)
    \ket{0}
    \right|^2
    =
    1 .
\label{eq:supp_completeness_amp}
\end{equation}
This relation is the completeness relation for the single-particle basis. Hence Eq.~\eqref{eq:supp_pair_sum_target} follows once one proves that the two transition probabilities within each symmetry-related pair are equal,
\begin{equation}
    \left|A_{j-\leftarrow m+}(t)\right|^2
    =
    \left|A_{j+\leftarrow m-}(t)\right|^2 .
\end{equation}

\SMsubsection{Symmetry-protected pairing}

For the \(\theta\)-quench, using \(C_{SC}=R_{SC}K\), one may choose a phase convention such that
\begin{equation}
    \eta_{j,+}^{\dagger}(t)
    =
    \sum_l
    s_l
    \left[
    v_{j,l}^{*}(t)a_l^\dagger
    +
    u_{j,l}^{*}(t)b_l^\dagger
    \right]
\label{eq:supp_theta_C_pair}
\end{equation}
when \(\eta_{j,-}^{\dagger}(t)\) is written as in Eq.~\eqref{eq:supp_measured_eta_minus}. Similarly,
\begin{equation}
    \widetilde{\eta}_{m,+}^{\dagger}(t)
    =
    \sum_l
    s_l
    \left[
    \widetilde v_{m-,l}^{*}(t)a_l^\dagger
    +
    \widetilde u_{m-,l}^{*}(t)b_l^\dagger
    \right].
\label{eq:supp_theta_C_tilde_pair}
\end{equation}
Substitution into Eq.~\eqref{eq:supp_transfer_amp_def} gives
\begin{equation}
      A_{j+\leftarrow m-}(t)
    =
    \sum_l
    s_l
    \left[
    v_{j,l}(t)\widetilde u_{m-,l}(t)
    +
    u_{j,l}(t)\widetilde v_{m-,l}(t)
    \right],
\end{equation}
while
\begin{equation}
      A_{j-\leftarrow m+}^{*}(t)
    =
    \sum_l
    s_l
    \left[
    u_{j,l}(t)\widetilde v_{m-,l}(t)
    +
    v_{j,l}(t)\widetilde u_{m-,l}(t)
    \right].
\end{equation}
Thus
\begin{equation}
      A_{j+\leftarrow m-}(t)
    =
      A_{j-\leftarrow m+}^{*}(t),
\end{equation}
and therefore
\begin{equation}
    \left|
      A_{j+\leftarrow m-}(t)
    \right|^2
    =
    \left|
      A_{j-\leftarrow m+}(t)
    \right|^2,
\label{eq:supp_prob_relation_theta}
\end{equation}

For the \(J_Y\)-quench, if
\begin{equation}
    \eta_{j,-}^{\dagger}(t)
    =
    \sum_{l=1}^{L}
    \left[
    u_{j,l}(t)a_l^\dagger
    +
    v_{j,l}(t)b_l^\dagger
    \right],
\label{eq:supp_JY_eta_minus_Cgauge}
\end{equation}
then the \(C_Y\)-paired mode can be chosen as
\begin{equation}
    \eta_{j,+}^{\dagger}(t)
    =
    \sum_{l=1}^{L}
    \left[
    u_{j,l}^{*}(t)a_l^\dagger
    -
    v_{j,l}^{*}(t)b_l^\dagger
    \right].
\label{eq:supp_JY_eta_plus_Cgauge}
\end{equation}
The evolved initial modes obey the same relation:
\begin{equation}
    \widetilde{\eta}_{m,-}^{\dagger}(t)
    =
    \sum_{l=1}^{L}
    \left[
    \widetilde u_{m-,l}(t)a_l^\dagger
    +
    \widetilde v_{m-,l}(t)b_l^\dagger
    \right],
\end{equation}
and
\begin{equation}
    \widetilde{\eta}_{m,+}^{\dagger}(t)
    =
    \sum_{l=1}^{L}
    \left[
    \widetilde u_{m-,l}^{*}(t)a_l^\dagger
    -
    \widetilde v_{m-,l}^{*}(t)b_l^\dagger
    \right].
\label{eq:supp_JY_tilde_eta_plus_Cgauge}
\end{equation}
Using Eqs.~\eqref{eq:supp_JY_eta_plus_Cgauge} and \eqref{eq:supp_JY_tilde_eta_plus_Cgauge}, the physical \(+\)-mode transfer amplitude is
\begin{equation}
      A_{j+\leftarrow m-}(t)
    =
    \sum_{l=1}^{L}
    \left[
    u_{j,l}(t)\widetilde u_{m-,l}(t)
    -
    v_{j,l}(t)\widetilde v_{m-,l}(t)
    \right].
\label{eq:supp_JY_A_jplus_mminus}
\end{equation}
On the other hand,
\begin{equation}
      A_{j-\leftarrow m+}^{*}(t)
    =
    \sum_{l=1}^{L}
    \left[
    u_{j,l}(t)\widetilde u_{m-,l}(t)
    -
    v_{j,l}(t)\widetilde v_{m-,l}(t)
    \right].
\label{eq:supp_JY_A_jminus_mplus_conj}
\end{equation}
Therefore,
\begin{equation}
      A_{j+\leftarrow m-}(t)
    =
      A_{j-\leftarrow m+}^{*}(t),
\end{equation}
and hence
\begin{equation}
    \left|
      A_{j+\leftarrow m-}(t)
    \right|^2
    =
    \left|
      A_{j-\leftarrow m+}(t)
    \right|^2 .
\label{eq:supp_JY_transfer_prob_relation}
\end{equation}

Using Eq.~\eqref{eq:supp_occ_transfer}, we have
\begin{equation}
    n_{j,-}(t)
    =
    \sum_m
    \left|
      A_{j-\leftarrow m-}(t)
    \right|^2 ,
\end{equation}
and
\begin{equation}
    n_{j,+}(t)
    =
    \sum_m
    \left|
      A_{j+\leftarrow m-}(t)
    \right|^2 .
\end{equation}
The symmetry relation gives
\begin{equation}
    n_{j,+}(t)
    =
    \sum_m
    \left|
      A_{j-\leftarrow m+}(t)
    \right|^2 .
\end{equation}
Therefore,
\begin{align}
    \langle\hat\rho_j(t)\rangle
    &=
    n_{j,-}(t)+n_{j,+}(t)
\nonumber\\
    &=
    \sum_m
    \left|
      A_{j-\leftarrow m-}(t)
    \right|^2
    +
    \sum_m
    \left|
      A_{j-\leftarrow m+}(t)
    \right|^2
\nonumber\\
    &=1,
\end{align}
where the final equality is the completeness identity Eq.~\eqref{eq:supp_missing_contribution}. This proves Eq.~\eqref{eq:supp_pair_sum_target}. For the zero modes, we have
\begin{equation}
    N_{\mathrm Z}(t)
    =
    \left\langle
    \eta_{1,+}^{\dagger}(t)\eta_{1,+}(t)
    +
    \eta_{1,-}^{\dagger}(t)\eta_{1,-}(t)
    \right\rangle
    =
    1 .
\end{equation}
Thus the quench can redistribute the occupation between the two members of the zero-mode pair, but the total occupation of the pair remains pinned to one in the half-filled state. To verify this relation numerically, we calculate \(N_{\mathrm Z}(t_f)=n_{1,+}(t_f)+n_{1,-}(t_f)\) for the three quench protocols. As shown in Fig.~\ref{fig:supp_NZproof}, \(N_{\mathrm Z}(t_f)\) remains equal to unity for all values of \(\tau_Q\), confirming Eq.~\eqref{eq:supp_pair_relationship}.

\begin{figure*}[t]
    \centering
    \includegraphics[width=1\textwidth]{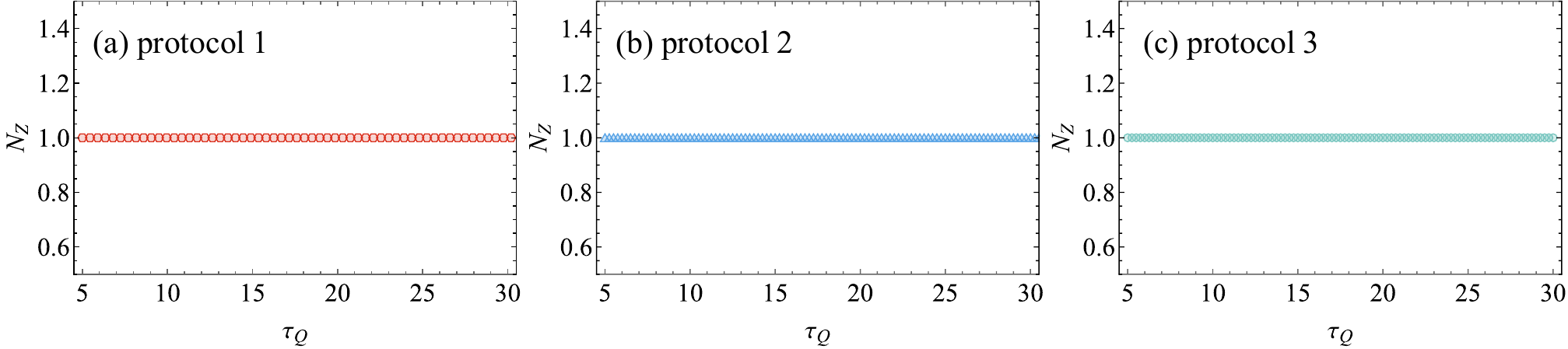}
    \caption{ Numerical verification of the zero-mode-pair occupation relation. The total occupation \(N_{\mathrm Z}(t_f)=n_{1,+}(t_f)+n_{1,-}(t_f)\) of the final zero-mode pair is plotted as a function of the quench time \(\tau_Q\) for (a) protocol 1, (b) protocol 2, and (c) protocol 3. In all three cases, \(N_{\mathrm Z}(t_f)\) remains pinned to unity within numerical precision, confirming Eq.~\eqref{eq:supp_pair_relationship}. }
    \label{fig:supp_NZproof}
\end{figure*}

\SMsection{Whittaker-Hill-type equation for protocol 1}
\label{sec:supp_theta_WH}

In this section, we derive the Whittaker-Hill-type equation for protocol 1. In this protocol, \(\mu=0\) is fixed and \(\theta\) is varied cyclically. The Bloch Hamiltonian is
\begin{equation}
    h_k(\theta)
    =
    d_{0,k}(\theta)I
    +
    d_{x,k}\sigma_x
    +
    d_{z,k}(\theta)\sigma_z ,
\label{eq:supp_theta_h}
\end{equation}
with
\begin{equation}
\left\{
\begin{aligned}
d_{0,k}(\theta)&=-2K\cos\theta\cos k,\\
d_{x,k}&=-2K\cos k,\\
d_{z,k}(\theta)&=2K\sin\theta\sin k .
\end{aligned}
\right.
\label{eq:supp_theta_d}
\end{equation}
The single-particle equation is
\begin{equation}
    i\frac{d}{dt}\Psi_k(t)=h_k(t)\Psi_k(t).
\end{equation}
The identity term \(d_{0,k}I\) contributes only a common phase. We remove it by writing
\begin{equation}
    \Psi_k(t)
    =
    e^{-i\zeta_k(t)}
    \widetilde{\Psi}_k(t),
    \qquad
    \zeta_k(t)=\int_{t_i}^{t}d_{0,k}(t')dt' .
\label{eq:supp_theta_common_phase}
\end{equation}
Then the reduced spinor satisfies
\begin{equation}
    i\partial_t\widetilde{\Psi}_k(t)
    =
    \left[
    d_{x,k}\sigma_x+d_{z,k}(t)\sigma_z
    \right]\widetilde{\Psi}_k(t).
\label{eq:supp_theta_reduced_eom}
\end{equation}
Writing
\begin{equation}
    \widetilde{\Psi}_k(t)
    =
    \begin{pmatrix}
    x_k(t)\\
    y_k(t)
    \end{pmatrix},
\end{equation}
one obtains
\begin{equation}
    i\dot x_k=d_{z,k}x_k+d_{x,k}y_k,
    \qquad
    i\dot y_k=d_{x,k}x_k-d_{z,k}y_k .
\label{eq:supp_theta_first_order}
\end{equation}
Eliminating \(y_k\) gives
\begin{equation}
    \ddot x_k+
    \left[
    d_{x,k}^2+d_{z,k}^2(\theta)+i\dot d_{z,k}(\theta)
    \right]x_k=0 .
\label{eq:supp_theta_second_time}
\end{equation}
Similarly, eliminating \(x_k\) gives
\begin{equation}
    \ddot y_k+
    \left[
    d_{x,k}^2+d_{z,k}^2(\theta)-i\dot d_{z,k}(\theta)
    \right]y_k=0 .
\label{eq:supp_theta_second_time_y}
\end{equation}

For a linear segment, we write
\begin{equation}
    \frac{d\theta}{dt}=s_\theta\frac{1}{\tau_s}.
\end{equation}
In the cyclic protocol, \(s_\theta=-1\) in both stages, with \(\tau_s=\tau_Q\) in the first stage and \(\tau_s=\tau_Q'=R\tau_Q\) in the second stage. Since
\begin{equation}
    \dot d_{z,k}
    =
    2K\sin k\cos\theta\,\dot\theta
    =
    s_\theta\frac{2K}{\tau_s}\sin k\cos\theta ,
\end{equation}
using \(\theta\) as the evolution variable gives
\begin{equation}
\begin{split}
\frac{d^2x_k}{d\theta^2}
+
\Big[
&4K^2\tau_s^2
\left(\cos^2k+\sin^2k\sin^2\theta\right)
\\
&+i\,s_\theta\,2K\tau_s\sin k\cos\theta
\Big]x_k=0 .
\end{split}
\label{eq:supp_theta_second_theta}
\end{equation}
For protocol 2, \(s_\theta=-1\), so the last term becomes
\(-i2K\tau_s\sin k\cos\theta\).

We now introduce the half-angle variable
\begin{equation}
    \varphi=\frac{\theta}{2},
    \qquad
    X_k(\varphi)=x_k(2\varphi).
\end{equation}
Since
\begin{equation}
    \frac{d^2x_k}{d\theta^2}
    =
    \frac14
    \frac{d^2X_k}{d\varphi^2},
\end{equation}
Eq.~\eqref{eq:supp_theta_second_theta} can be written as
\begin{equation}
    X_k''(\varphi)
    +
    \left[
      A_k+  B_k\cos2\varphi+  C_k\cos4\varphi
    \right]X_k(\varphi)=0 .
\label{eq:supp_theta_WH}
\end{equation}
The coefficients are
\begin{equation}
\left\{
\begin{aligned}
      A_k&=8K^2\tau_s^2(1+\cos^2 k),\\
      B_k&=8i\,s_\theta K\tau_s\sin k,\\
      C_k&=-8K^2\tau_s^2\sin^2 k .
\end{aligned}
\right.
\label{eq:supp_theta_WH_coeff_general}
\end{equation}
For protocol 2, \(s_\theta=-1\), and therefore
\begin{equation}
      B_k=-8iK\tau_s\sin k .
\end{equation}
Equation~\eqref{eq:supp_theta_WH} is a Whittaker-Hill-type equation. It is exact for the reduced two-level equation after removing the common phase \(d_{0,k}I\).

We now extract the interference phase. In the slow-quench regime, the leading term in Eq.~\eqref{eq:supp_theta_second_theta} is the term proportional to \(\tau_s^2\). Therefore, in the adiabatic region,
\begin{equation}
x_k(\theta)
\approx
\frac{1}{\sqrt{\Lambda_k(\theta)}}
\left[
A_k e^{i\Phi_k(\theta)}
+
B_k e^{-i\Phi_k(\theta)}
\right],
\label{eq:supp_theta_wkb}
\end{equation}
where
\begin{equation}
    \Phi_k(\theta)
    =
    2K\tau_s\int_{\theta_0}^{\theta}
    \Lambda_k(\theta')d\theta' .
\label{eq:supp_theta_Phi_def}
\end{equation}
The relative phase between the two WKB branches in one segment is thus
\begin{equation}
    2\Phi_k
    =
    4K\tau_s\int d\theta\,\Lambda_k(\theta)
    =
    \tau_s\int d\theta\,\Delta_{\mathrm B}(k,\theta).
\label{eq:supp_theta_segment_phase}
\end{equation}
The branch splitting of the reduced two-level problem is
\begin{equation}
    \Delta_{\mathrm B}(k,\theta)
    =
    4K\Lambda_k(\theta),
\label{eq:supp_theta_branch_splitting}
\end{equation}
where
\begin{equation}
    \Lambda_k(\theta)
    =
    \sqrt{
    \cos^2 k+\sin^2\theta\,\sin^2 k
    }.
\label{eq:supp_theta_Lambda}
\end{equation}
Here \(\Delta_{\mathrm B}(k,\theta)\) is the branch splitting at fixed \(k\). The coefficients \(A_k\) and \(B_k\) are determined by the initial condition in the outer adiabatic region. Let
\begin{equation}
    \Lambda_{k0}\equiv\Lambda_k(\theta_0),
    \qquad
    \Phi_k(\theta_0)=0 .
\end{equation}
At \(\theta=\theta_0\),
\begin{equation}
    x_k(\theta_0)=\frac{A_k+B_k}{\sqrt{\Lambda_{k0}}}.
\end{equation}
If the initial state is the instantaneous lower-band eigenstate, its first component may be chosen as
\begin{equation}
    x_{k0}
    =
    \frac{\cos k}
    {\sqrt{2\Lambda_{k0}\left(\Lambda_{k0}+\sin\theta_0\sin k\right)}} .
\label{eq:supp_theta_x0}
\end{equation}
Hence,
\begin{equation}
    A_k+B_k=\sqrt{\Lambda_{k0}}\,x_{k0}.
\label{eq:supp_theta_AplusB}
\end{equation}
To fix \(A_k-B_k\), we use the first-order equation
\begin{equation}
    i\dot x_k=d_{z,k}x_k+d_{x,k}y_k .
\end{equation}
For the instantaneous lower-band eigenstate, this gives
\begin{equation}
    x_k'(\theta_0)
    =
    -i\,2K\tau_s\,\Lambda_{k0}\,x_{k0}.
\label{eq:supp_theta_xprime0}
\end{equation}
On the other hand, differentiating Eq.~\eqref{eq:supp_theta_wkb} gives
\begin{equation}
    x_k'(\theta_0)
    =
    Q_0'(A_k+B_k)
    +
    iQ_0\Phi_k'(\theta_0)(A_k-B_k),
\end{equation}
where
\begin{equation}
    Q_0=\Lambda_{k0}^{-1/2},
    \qquad
    \Phi_k'(\theta_0)=2K\tau_s\Lambda_{k0}.
\end{equation}
Since
\begin{equation}
    Q_0'
    =
    -\frac{\sin^2k\sin2\theta_0}{4\Lambda_{k0}^{5/2}},
\end{equation}
we obtain
\begin{equation}
    A_k-B_k
    =
    -\sqrt{\Lambda_{k0}}\,x_{k0}
    -
    \frac{i\sin^2k\sin2\theta_0}
    {8K\tau_s\Lambda_{k0}^{5/2}}
    x_{k0}.
\label{eq:supp_theta_AminusB}
\end{equation}
Therefore,
\begin{equation}
    A_k
    =
    -\frac{i\sin^2k\sin2\theta_0}
    {16K\tau_s\Lambda_{k0}^{5/2}}
    x_{k0},
\label{eq:supp_theta_A_coeff}
\end{equation}
and
\begin{equation}
    B_k
    =
    \sqrt{\Lambda_{k0}}\,x_{k0}
    +
    \frac{i\sin^2k\sin2\theta_0}
    {16K\tau_s\Lambda_{k0}^{5/2}}
    x_{k0}.
\label{eq:supp_theta_B_coeff}
\end{equation}
These coefficients determine the relative weights of the two WKB branches in the adiabatic region. However, the WKB approximation is not used here to compute the Kibble-Zurek scaling of the excitation probability. The nonadiabatic amplitudes are generated in narrow regions around the critical points, where the WKB condition is not reliable. The role of WKB in the present work is to extract the propagation phase accumulated after the two-branch superposition has been created.

We now determine the integration interval for protocol 2. The cyclic protocol is
\begin{equation}
\theta(t)=
\begin{cases}
\theta_m-\dfrac{t}{\tau_Q},
& t_i\le t\le0,\\[1.0ex]
\theta_m-\dfrac{t}{\tau_Q'},
& 0\le t\le t_f,
\end{cases}
\qquad
\tau_Q'=R\tau_Q .
\label{eq:supp_theta_cyclic_protocol}
\end{equation}
The conditions
\begin{equation}
    \theta(t_i)=\frac{\pi}{2},
    \qquad
    \theta(t_f)=-\frac{3\pi}{2}
\end{equation}
give
\begin{equation}
    t_i=\tau_Q\left(\theta_m-\frac{\pi}{2}\right),
    \qquad
    t_f=\tau_Q'\left(\theta_m+\frac{3\pi}{2}\right).
\label{eq:supp_theta_titf}
\end{equation}
We choose
\begin{equation}
    -\pi<\theta_m<0,
\end{equation}
so that the first critical point \(\theta_c^{(1)}=0\) lies in the first stage, while the second critical point \(\theta_c^{(2)}=-\pi\) lies in the second stage. Their times are
\begin{equation}
    t_c^{(1)}=\tau_Q\theta_m,
    \qquad
    t_c^{(2)}=\tau_Q'(\theta_m+\pi).
\label{eq:supp_theta_critical_times}
\end{equation}

Before the first critical dynamics, the state follows a single adiabatic branch. The phase accumulated in this part is common to the two histories that later interfere and therefore cancels from the relative phase. After the first critical dynamics, the local band mixing creates a superposition of the two adiabatic branches. The interference phase is therefore the relative propagation phase accumulated from \(\theta=0\) to \(\theta=-\pi\), with the turning point \(\theta_m\) separating the two linear segments.

Thus the relevant phase can be written as
\begin{equation}
\begin{split}
\Delta\varphi_{\mathrm B}(k)
&=
4K
\left[
\int_{t_c^{(1)}}^{0}
\Lambda_k[\theta(t)]dt
+
\int_{0}^{t_c^{(2)}}
\Lambda_k[\theta(t)]dt
\right]
+\delta_k .
\end{split}
\label{eq:supp_theta_phase_time}
\end{equation}
The constant \(\delta_k\) contains local phase contributions from the critical regions and does not change the leading period. Converting the time integrals into \(\theta\)-integrals gives
\begin{equation}
\begin{split}
\int_{t_c^{(1)}}^{0}
\Lambda_k[\theta(t)]dt
&=
\tau_Q\int_{\theta_m}^{0}
\Lambda_k(\theta)d\theta ,
\\
\int_{0}^{t_c^{(2)}}
\Lambda_k[\theta(t)]dt
&=
R\tau_Q\int_{-\pi}^{\theta_m}
\Lambda_k(\theta)d\theta .
\end{split}
\label{eq:supp_theta_time_to_theta}
\end{equation}
Therefore,
\begin{equation}
\begin{split}
\Delta\varphi_{\mathrm B}(k)
&=
4K\tau_Q
\left[
\int_{\theta_m}^{0}\Lambda_k(\theta)d\theta
+
R\int_{-\pi}^{\theta_m}\Lambda_k(\theta)d\theta
\right]
+\delta_k
\\
&=
\tau_Q
\int_{\theta_m}^{0}
\Delta_{\mathrm B}(k,\theta)d\theta
+
R\tau_Q
\int_{-\pi}^{\theta_m}
\Delta_{\mathrm B}(k,\theta)d\theta
+\delta_k .
\end{split}
\label{eq:supp_theta_phase}
\end{equation}

We now express the integrals in Eq.~\eqref{eq:supp_theta_phase} in terms of elliptic integrals. From Eq.~\eqref{eq:supp_theta_Lambda},
\begin{equation}
\Lambda_k(\theta)
=
\sqrt{1-\sin^2 k\,\cos^2\theta}.
\end{equation}
Let
\begin{equation}
    m_k=\sin^2 k,
    \qquad
    u=\theta+\frac{\pi}{2}.
\end{equation}
Then
\begin{equation}
    \Lambda_k(\theta)
    =
    \sqrt{1-m_k\sin^2u},
\end{equation}
and
\begin{equation}
    \int \Lambda_k(\theta)d\theta
    =
    E(u|m_k),
\label{eq:supp_theta_elliptic_indef}
\end{equation}
where \(E(u|m_k)\) is the incomplete elliptic integral of the second kind.

Using
\begin{equation}
    E\!\left(\frac{\pi}{2}\middle|m_k\right)=E(m_k),
    \qquad
    E\!\left(-\frac{\pi}{2}\middle|m_k\right)=-E(m_k),
\end{equation}
we obtain
\begin{equation}
\int_{\theta_m}^{0}\Lambda_k(\theta)d\theta
=
E(m_k)
-
E\!\left(\theta_m+\frac{\pi}{2}\middle|m_k\right),
\label{eq:supp_theta_int1}
\end{equation}
and
\begin{equation}
\int_{-\pi}^{\theta_m}\Lambda_k(\theta)d\theta
=
E(m_k)
+
E\!\left(\theta_m+\frac{\pi}{2}\middle|m_k\right).
\label{eq:supp_theta_int2}
\end{equation}
Substituting Eqs.~\eqref{eq:supp_theta_int1} and \eqref{eq:supp_theta_int2} into Eq.~\eqref{eq:supp_theta_phase}, the leading dynamical part of the interference phase becomes
\begin{equation}
\Delta\varphi_{\mathrm B}^{\rm dyn}(k)
=
4K\tau_Q
\left[
(1+R)E(m_k)
+
(R-1)
E\!\left(\theta_m+\frac{\pi}{2}\middle|m_k\right)
\right].
\label{eq:supp_theta_phase_elliptic}
\end{equation}

In the slow-quench regime, the dominant contribution comes from the critical mode
\begin{equation}
    k^\ast=\frac{\pi}{2}.
\end{equation}
For this mode,
\begin{equation}
    m_{k^\ast}=1,
    \qquad
    E(1)=1.
\end{equation}
Since \(-\pi<\theta_m<0\),
\begin{equation}
    -\frac{\pi}{2}
    <
    \theta_m+\frac{\pi}{2}
    <
    \frac{\pi}{2}.
\end{equation}
In this interval,
\begin{equation}
E\!\left(\theta_m+\frac{\pi}{2}\middle|1\right)
=
\int_0^{\theta_m+\pi/2}\cos u\,du
=
\cos\theta_m .
\label{eq:supp_theta_E_m1}
\end{equation}
Therefore,
\begin{equation}
\Delta\varphi_{\mathrm B}(k^\ast)
\simeq
4K\tau_Q
\left[
(1+R)+(R-1)\cos\theta_m
\right]
+
\delta_{k^\ast}.
\label{eq:supp_theta_phase_kstar}
\end{equation}
The period is determined by
\begin{equation}
\Delta\varphi_{\mathrm B}(k^\ast,\tau_Q+T_{\mathrm B})
-
\Delta\varphi_{\mathrm B}(k^\ast,\tau_Q)
=
2\pi ,
\end{equation}
which gives
\begin{equation}
    T_{\mathrm B}
    =
    \frac{\pi}
    {2K\left[(1+R)+(R-1)\cos\theta_m\right]} .
\label{eq:supp_theta_TB}
\end{equation}
For \(R=1\), this reduces to
\begin{equation}
    T_{\mathrm B}=\frac{\pi}{4K}.
\end{equation}

Finally, we note how this bulk phase result is connected to the doubled zero-mode period discussed in the main text. The bulk interference compares the two ordinary adiabatic branches, whose branch splitting is \(4K\Lambda_k(\theta)\). The zero-mode-related interference compares the occupied zero mode with one bulk-like branch in the OBC spectrum, giving an effective energy scale approximately equal to one half of the ordinary bulk particle-hole energy gap. Therefore, the corresponding phase satisfies
\begin{equation}
    \Delta\varphi_{\mathrm Z}
    \simeq
    \frac{1}{2}\Delta\varphi_{\mathrm B}
    +
    \mathrm{const.},
\end{equation}
and hence
\begin{equation}
    T_{\mathrm Z}\simeq 2T_{\mathrm B}.
\end{equation}

\SMsection{Weber-type equation and WKB phase for protocol 3}
\label{sec:supp_JY_weber}

In this section, we derive the Weber-type equation for protocol 3 and extract the WKB phase used in the main text. In protocol 3 \(\theta=\pi/2\) is fixed and \(\mu\) is varied. The Bloch Hamiltonian is
\begin{equation}
    h_k(\mu)
    =
    -2K(\mu+\cos k)\sigma_x
    +
    2K\sin k\,\sigma_z .
\label{eq:supp_JY_h_original}
\end{equation}
This Hamiltonian is traceless. To put the time-dependent term on \(\sigma_z\), we perform the time-independent rotation
\begin{equation}
    \widetilde{\Psi}_k(t)
    =
    U_y\Psi_k(t),
    \qquad
    U_y=e^{-i\pi\sigma_y/4}.
\label{eq:supp_JY_rotation}
\end{equation}
Since \(U_y\) is independent of time, this rotation does not generate an additional gauge term. The rotated spinor satisfies
\begin{equation}
    i\partial_t\widetilde{\Psi}_k(t)
    =
    \widetilde h_k(\mu)\widetilde{\Psi}_k(t),
\end{equation}
where
\begin{equation}
    \widetilde h_k(\mu)
    =
    U_y h_k(\mu)U_y^\dagger
    =
    2K(\mu+\cos k)\sigma_z
    +
    2K\sin k\,\sigma_x .
\label{eq:supp_JY_rotated_h}
\end{equation}
Writing
\begin{equation}
    \widetilde{\Psi}_k(t)
    =
    \begin{pmatrix}
    u_k(t)\\
    v_k(t)
    \end{pmatrix},
\end{equation}
one obtains
\begin{equation}
    i\dot u_k=a_k(\mu)u_k+b_kv_k,
    \qquad
    i\dot v_k=b_ku_k-a_k(\mu)v_k,
\label{eq:supp_JY_first_order}
\end{equation}
with
\begin{equation}
    a_k(\mu)=2K(\mu+\cos k),
    \qquad
    b_k=2K\sin k .
\label{eq:supp_JY_ab_def}
\end{equation}
Eliminating \(v_k\) gives
\begin{equation}
    \ddot u_k+
    \left[
    a_k^2(\mu)+b_k^2+i\dot a_k(\mu)
    \right]u_k=0 .
\label{eq:supp_JY_second_time}
\end{equation}

For each linear segment of protocol 3, we write
\begin{equation}
    \frac{d\mu}{dt}=s_\mu\frac{1}{\tau_s},
    \qquad
    s_\mu=\pm1.
\end{equation}
Here \(\tau_s=\tau_Q\) in the first stage and \(\tau_s=\tau_Q'=R\tau_Q\) in the second stage. Using \(\mu\) as the evolution variable, Eq.~\eqref{eq:supp_JY_second_time} becomes
\begin{equation}
    \frac{d^2u_k}{d\mu^2}
    +
    \left[
    \tau_s^2\varepsilon_k^2(\mu)
    +
    i\,s_\mu\,2K\tau_s
    \right]u_k=0,
\label{eq:supp_JY_second_mu}
\end{equation}
where
\begin{equation}
    \varepsilon_k(\mu)
    =
    \sqrt{a_k^2(\mu)+b_k^2}
    =
    2K\sqrt{(\mu+\cos k)^2+\sin^2 k}.
\label{eq:supp_JY_epsilon}
\end{equation}
The last term in Eq.~\eqref{eq:supp_JY_second_mu} depends on the sweep direction. It changes the local connection phase, but not the leading \(\tau_Q\)-linear propagation phase that determines the interference period.

To show that Eq.~\eqref{eq:supp_JY_second_mu} is a Weber-type equation, we introduce
\begin{equation}
    \xi=\mu+\cos k .
\end{equation}
Then
\begin{equation}
    \frac{d^2u_k}{d\xi^2}
    +
    \left[
    4K^2\tau_s^2(\xi^2+\sin^2 k)
    +
    i\,s_\mu\,2K\tau_s
    \right]u_k=0 .
\label{eq:supp_JY_xi}
\end{equation}
With
\begin{equation}
    z=e^{i\pi/4}\sqrt{4K\tau_s}\,\xi ,
\end{equation}
Eq.~\eqref{eq:supp_JY_xi} becomes
\begin{equation}
    \frac{d^2u_k}{dz^2}
    +
    \left[
    \nu_k+\frac12-\frac{z^2}{4}
    \right]u_k=0,
\label{eq:supp_JY_parabolic}
\end{equation}
where
\begin{equation}
    \nu_k
    =
    \frac{s_\mu-1}{2}
    -
    iK\tau_s\sin^2 k .
\label{eq:supp_JY_nu}
\end{equation}
This is the standard parabolic-cylinder equation, namely the Weber equation.

We next fix the integration interval that enters the interference phase. Protocol 3 is
\begin{equation}
\mu(t)=
\begin{cases}
\mu_m+\dfrac{t}{\tau_Q}, & t_i\le t\le0,\\[1.0ex]
\mu_m-\dfrac{t}{\tau_Q'}, & 0\le t\le t_f,
\end{cases}
\qquad
\tau_Q'=R\tau_Q .
\label{eq:supp_JY_protocol}
\end{equation}
The endpoints are determined by \(\mu(t_i)=0\) and \(\mu(t_f)=0\), giving
\begin{equation}
    t_i=-\mu_m\tau_Q,
    \qquad
    t_f=\mu_m\tau_Q' .
\end{equation}
The two critical dynamics occur at \(\mu=1\). Therefore,
\begin{equation}
    t_c^{(1)}=(1-\mu_m)\tau_Q,
    \qquad
    t_c^{(2)}=(\mu_m-1)\tau_Q' .
\label{eq:supp_JY_critical_times}
\end{equation}
Before the first critical dynamics, the system remains on a single adiabatic branch. This part only contributes a common phase and therefore does not enter the relative phase responsible for interference. The phase relevant to the interference period is accumulated between the two critical dynamics, namely from \(t_c^{(1)}\) to \(t_c^{(2)}\).

In the slow-quench regime, the dominant term in Eq.~\eqref{eq:supp_JY_second_mu} is the term proportional to \(\tau_s^2\). The WKB solution in the adiabatic region has the form
\begin{equation}
u_k(\mu)
\approx
\frac{1}{\sqrt{\varepsilon_k(\mu)}}
\left[
A_k e^{i\tau_s\int^\mu \varepsilon_k(\mu')d\mu'}
+
B_k e^{-i\tau_s\int^\mu \varepsilon_k(\mu')d\mu'}
\right].
\label{eq:supp_JY_wkb}
\end{equation}
The relative phase between the two WKB branches in one segment is
\begin{equation}
    2\tau_s\int d\mu\,\varepsilon_k(\mu)
    =
    \tau_s\int d\mu\,\Delta_{\mathrm B}(k,\mu),
\end{equation}
where the branch splitting is
\begin{equation}
    \Delta_{\mathrm B}(k,\mu)
    =
    2\varepsilon_k(\mu)
    =
    4K\sqrt{(\mu+\cos k)^2+\sin^2 k}.
\label{eq:supp_JY_branch_splitting}
\end{equation}
Here \(\Delta_{\mathrm B}(k,\mu)\) is the branch splitting at fixed \(k\), not yet the bulk energy gap. 

We first clarify a general point used in both protocols.  Consider a fixed
momentum \(k\) after the common identity term has been removed.  Let
\(t_c^{(1)}\) and \(t_c^{(2)}\) denote the two critical passages.  Outside
small neighborhoods of these passages, the dynamics is described by
instantaneous branches \(|+\rangle\) and \(|-\rangle\).  Suppose the first
critical passage creates amplitudes \(r_{k,+}^{(1)}\) and \(r_{k,-}^{(1)}\)
on the two branches, and the second critical passage recombines them into the
final branch \(\alpha=\pm\) with amplitudes \(r_{k,\alpha +}^{(2)}\) and
\(r_{k,\alpha -}^{(2)}\).  Then the final amplitude can be written as
\begin{equation}
  A_{k,\alpha}(t_f)
=
e^{-i\phi_{k,\alpha}^{\rm out}}
\left[
r_{k,\alpha +}^{(2)} r_{k,+}^{(1)} e^{-i\Phi_{k,+}}
+
r_{k,\alpha -}^{(2)} r_{k,-}^{(1)} e^{-i\Phi_{k,-}}
\right]
e^{-i\phi_k^{\rm in}},
\label{eq:supp_general_two_passage_amp}
\end{equation}
where
\begin{equation}
\Phi_{k,\pm}
=
\int_{t_c^{(1)}}^{t_c^{(2)}}
E_{k,\pm}(t)\,dt .
\label{eq:supp_general_branch_phase}
\end{equation}
Here \(\phi_k^{\rm in}\) is the phase accumulated before the first critical
passage, and \(\phi_{k,\alpha}^{\rm out}\) is the phase accumulated after the
second critical passage along the final branch \(\alpha\).

The final occupation of branch \(\alpha\) is
\begin{equation}
p_{k,\alpha}
=
|  A_{k,\alpha}(t_f)|^2 .
\end{equation}
Substituting Eq.~\eqref{eq:supp_general_two_passage_amp}, the input and output
phases cancel, giving
\begin{equation}
\begin{split}
p_{k,\alpha}
={}&
\left|r_{k,\alpha +}^{(2)} r_{k,+}^{(1)}\right|^2
+
\left|r_{k,\alpha -}^{(2)} r_{k,-}^{(1)}\right|^2
\\
&+
2\,{\rm Re}
\left[
r_{k,\alpha +}^{(2)} r_{k,+}^{(1)}
\left(r_{k,\alpha -}^{(2)} r_{k,-}^{(1)}\right)^*
e^{-i\Delta\Phi_k}
\right],
\end{split}
\label{eq:supp_general_two_passage_prob}
\end{equation}
where the relative phase is
\begin{equation}
\Delta\Phi_k
=
\Phi_{k,+}-\Phi_{k,-}
=
\int_{t_c^{(1)}}^{t_c^{(2)}}
\left[E_{k,+}(t)-E_{k,-}(t)\right]dt .
\label{eq:supp_general_relative_phase}
\end{equation}
Thus only the phase accumulated between the two critical passages determines
the leading oscillatory period of a final occupation.  The phase before
\(t_c^{(1)}\) is common because only one adiabatic branch has been followed,
whereas the phase after \(t_c^{(2)}\) is an output phase and cancels in the
occupation probability.  This is the same phase-window structure as in
Landau-Zener-Stueckelberg interferometry~\cite{Shevchenko2010LZS}.

Using Eq.~\eqref{eq:supp_JY_protocol}, the time integral between the two critical dynamics can be written as
\begin{equation}
\begin{split}
\Delta\varphi_{\mathrm B}(k)
&=
\int_{t_c^{(1)}}^{0}
\Delta_{\mathrm B}[k,\mu(t)]\,dt
+
\int_{0}^{t_c^{(2)}}
\Delta_{\mathrm B}[k,\mu(t)]\,dt
+\delta_k
\\
&=
\tau_Q\int_{1}^{\mu_m}
\Delta_{\mathrm B}(k,\mu)d\mu
+
\tau_Q'\int_{1}^{\mu_m}
\Delta_{\mathrm B}(k,\mu)d\mu
+\delta_k
\\
&=
(1+R)\tau_Q
\int_{1}^{\mu_m}
\Delta_{\mathrm B}(k,\mu)d\mu
+\delta_k .
\end{split}
\label{eq:supp_JY_phase}
\end{equation}
The constant \(\delta_k\) contains the local phase contribution from the critical regions and does not change the leading period. It is useful to clarify why the phase interval in
Eq.~\eqref{eq:supp_JY_phase} is restricted to the region between the two
critical dynamics. 

The dominant critical mode is \(k^\ast=\pi\). In the interval \(1<\mu<\mu_m\),
\begin{equation}
    \Delta_{\mathrm B}(k^\ast,\mu)
    =
    4K(\mu-1).
\end{equation}
Thus
\begin{equation}
\begin{split}
    \Delta\varphi_{\mathrm B}(k^\ast)
    &=
    (1+R)\tau_Q
    \int_{1}^{\mu_m}4K(\mu-1)d\mu
    +
    \delta_{k^\ast}
    \\
    &=
    2K(1+R)(\mu_m-1)^2\tau_Q
    +
    \delta_{k^\ast}.
\end{split}
\label{eq:supp_JY_phase_kstar}
\end{equation}
The period is determined by
\begin{equation}
    \Delta\varphi_{\mathrm B}(k^\ast,\tau_Q+T_{\mathrm B})
    -
    \Delta\varphi_{\mathrm B}(k^\ast,\tau_Q)
    =
    2\pi ,
\end{equation}
which gives
\begin{equation}
    T_{\mathrm B}
    =
    \frac{\pi}{K(1+R)(\mu_m-1)^2}.
\label{eq:supp_JY_TB}
\end{equation}

\endgroup

\end{document}